\documentclass[preprint]{aastex}

\usepackage{emulateapj5}
\usepackage{epsf}
\usepackage{psfig}
\usepackage{graphicx}
\usepackage{lscape}

\begin{document}
\title{Very Low-Mass Stellar and Substellar Companions to Solar-Like Stars from MARVELS I: A Low Mass Ratio Stellar Companion to TYC 4110-01037-1 in a 79-day Orbit}
\author{John P. Wisniewski\altaffilmark{1}, Jian Ge\altaffilmark{2}, Justin R. Crepp\altaffilmark{21}, Nathan De Lee\altaffilmark{2,3}, Jason Eastman\altaffilmark{4,5,6}, Massimiliano Esposito\altaffilmark{7,8}, Scott W. Fleming\altaffilmark{2,9},B. Scott Gaudi\altaffilmark{4}, Luan Ghezzi\altaffilmark{10,11}, Jonay I. Gonzalez Hernandez\altaffilmark{7,8}, Brian L. Lee\altaffilmark{2}, Keivan G. Stassun\altaffilmark{3,12,13}, Eric Agol\altaffilmark{1}, Carlos Allende Prieto\altaffilmark{7,8}, Rory Barnes\altaffilmark{1},  Dmitry Bizyaev\altaffilmark{14}, Phillip Cargile\altaffilmark{3}, Liang Chang\altaffilmark{2}, Luiz N. Da Costa\altaffilmark{10,11}, G.F. Porto De Mello\altaffilmark{10,15}, Bruno Femen\'{\i}a\altaffilmark{7,8}, Leticia D. Ferreira\altaffilmark{10,15}, Bruce Gary\altaffilmark{3}, Leslie Hebb\altaffilmark{3}, Jon Holtzman\altaffilmark{16}, Jian Liu\altaffilmark{2}, Bo Ma\altaffilmark{2}, Claude E. Mack III\altaffilmark{3}, Suvrath Mahadevan\altaffilmark{9,17}, Marcio A.G. Maia\altaffilmark{10,11}, Duy Cuong Nguyen\altaffilmark{2}, Ricardo L.C. Ogando\altaffilmark{10,11}, Daniel J. Oravetz\altaffilmark{14}, Martin Paegert\altaffilmark{3}, Kaike Pan\altaffilmark{14}, Joshua Pepper\altaffilmark{3}, Rafael Rebolo \altaffilmark{7,8,18}, Basilio Santiago\altaffilmark{10,19}, Donald P. Schneider\altaffilmark{9,17}, Alaina C Shelden\altaffilmark{14}, Audrey Simmons\altaffilmark{14}, Benjamin M. Tofflemire\altaffilmark{1,20}, Xiaoke Wan\altaffilmark{2}, Ji Wang\altaffilmark{2}, Bo Zhao\altaffilmark{2}}

\altaffiltext{1}{Astronomy Department, University of Washington, Box 351580, Seattle, WA 98195, USA; jwisnie@u.washington.edu}
\altaffiltext{2}{Department of Astronomy, University of Florida, 211 Bryant Space Science Center, Gainesville, FL, 32611-2055, USA}
\altaffiltext{3}{Department of Physics and Astronomy, Vanderbilt University, Nashville, TN 37235, USA}
\altaffiltext{4}{Department of Astronomy, The Ohio State University, 140 West 18th Avenue, Columbus, OH 43210, USA}
\altaffiltext{5}{Las Cumbres Observatory Global Telescope Network, 6740 Cortona Drive, Suite 102, Santa Barbara, CA 93117, USA}
\altaffiltext{6}{Department of Physics Broida Hall, University of California, Santa Barbara, CA 93106, USA}
\altaffiltext{7}{Instituto de Astrof\'{\i}sica de Canarias (IAC), E-38205 La Laguna, Tenerife, Spain}
\altaffiltext{8}{Departamento de Astrof\'{\i}sica, Universidad de La Laguna, 38206 La Laguna, Tenerife, Spain}
\altaffiltext{9}{Department of Astronomy and Astrophysics, The Pennsylvania State University, 525 Davey Laboratory, University Park, PA 16802, USA}
\altaffiltext{10}{Laboratório Interinstitucional de e-Astronomia-LIneA, Rio de Janeiro, RJ 20921-400, Brazil}
\altaffiltext{11}{Observatório Nacional, Rio de Janeiro, RJ 20921-400, Brazil }
\altaffiltext{12}{Department of Physics, Fisk University, Nashville, TN, USA}
\altaffiltext{13}{Department of Physics, Massachusetts Institute of Technology, Cambridge, MA 02139}
\altaffiltext{14}{Apache Point Observatory, P.O. Box 59, Sunspot, NM 88349-0059, USA}
\altaffiltext{15}{Universidade Federal do Rio de Janeiro, Observatório do Valongo, Ladeira do Pedro Antonio 43, 20080-090 Rio de Janeiro, Brazil}
\altaffiltext{16}{Department of Astronomy, New Mexico State University, Box 30001, Las Cruces, NM 880033, USA}
\altaffiltext{17}{Center for Exoplanets and Habitable Worlds, Pennsylvania State University, University Park, PA 16802, USA}
\altaffiltext{18}{Consejo Superior de Investigaciones Cient\'{\i}ficas, Spain}
\altaffiltext{19}{Instituto de Fisica, UFRGS, Porto Alegre, RS 91501-970, Brazil}
\altaffiltext{20}{Astronomy Department, University of Wisconsin-Madison, 475 N Charter St, Madison, WI 53706, USA}
\altaffiltext{21}{California Institute of Technology, Department of Astrophysics, 1200 E. California Blvd., Pasadena, CA 91125, USA}

\begin{abstract} 

TYC 4110-01037-1 has a low-mass stellar companion, whose small mass ratio and short orbital period are atypical amongst solar-like ($T_{eff} \lesssim 6000$ K) 
binary systems.  Our analysis of TYC 4110-01037-1 reveals it to be 
a moderately aged ($\lesssim$5 Gyr) solar-like star having a mass of 1.07 $\pm$0.08 M$_{\Sun}$ and radius of 0.99 $\pm$0.18 R$_{\Sun}$.  
We analyze 32 radial velocity measurements from the SDSS-III MARVELS survey as well as 6 supporting radial velocity measurements 
from the SARG spectrograph on the 3.6m TNG telescope obtained 
over a period of $\sim$2 years.  The best Keplerian orbital fit parameters were found to have a period of 78.994 $\pm$0.012 days, an eccentricity 
of 0.1095$\pm$0.0023, and a semi-amplitude of 4199$\pm$11 m s$^{-1}$.  We determine the minimum companion mass (if $\sin i$ = 1) to be 97.7 $\pm$5.8 M$_{Jup}$.  
The system's companion to host star mass ratio, $\geq$0.087 $\pm$0.003, places it at the lowest end of observed values for  
short period stellar companions to solar-like ($T_{eff} \lesssim 6000$ K) stars.  One possible way to create such a system would be if a triple-component stellar multiple broke up into a short period, low $q$ binary during the cluster dispersal phase of its lifetime.   A candidate tertiary body has been identified in the system via single-epoch, high contrast imagery.  If this object is confirmed to be co-moving, we estimate it would be a dM4 star.  We present these results in the context of our larger-scale effort to constrain the statistics of 
low mass stellar and brown dwarf companions to FGK-type stars via the MARVELS survey. 
\end{abstract}


\section{Introduction} \label{intro}

Observations of the mass distribution of substellar and low-mass stars as a function of orbital separation, using a variety of techniques, provide key constraints which 
influence our understanding of the process of planetary and star formation \citep{bur07,sta09,kra11,sah11}.  Early results from the Kepler transit search program, which is sensitive to short to moderate period companions, show the size distribution of candidate companions increases towards smaller planets, reaching a maximum at a few R$_{\earth}$ \citep{bo11a,bo11b}.  Numerous large radial velocity surveys are constraining the prevalence of Jovian-mass giant planets at close and intermediate orbital separations (e.g. California \& Carnegie teams, \citealt{mar00}; AAPS, \citealt{tin01}; CORALIE, \citealt{udr00}; HARPS, \citealt{may04}) and the prevalence of brown dwarf (BD) and very low mass stellar companions at close and intermediate separations \citep{mar00,may01,vog02,pat07,lee11}.  At much larger orbital separations, high contrast imaging surveys inform our understanding of the mass distribution of planets \citep{kal08,mar08,mar10,lag10} and BDs \citep{tha09,bil10,jan11,wah11}, while the mass distribution of very low mass stellar companions has been explored by 
high contrast imaging \citep{kra11}, spectroscopic \citep{duq91}, and interferometric \citep{rag10} studies.

Many of the results from these surveys have found interesting, and at times conflicting, trends motivating subsequent programs to investigate their origin.  
In the BD regime, a deficit of short period ($a$ $\leq$ 5 AU) companions to solar-type primary stars has long been designated the apparent ``brown dwarf desert'' 
\citep{mar00}.  Investigations of the BD frequency at larger orbital separations (e.g. \citealt{giz01,met04,mcc04,met09,kra11}) have led to conflicting assertions as to whether
the BD desert extends to larger orbital separations.  Multiplicity studies of  low-mass stellar companions to solar-like stars have found evidence of a unimodal period 
distribution with peak periods ranging from $\sim$180 years \citep{duq91} to $\sim$300 years \citep{rag10}.  Interestingly, recent work by \citet{met09} has presented 
tentative evidence that the companion mass function of brown dwarf and low-mass stellar companions around solar-like stars could be represented by a universal function.  

At the short-period ($P$ $<$100 days) tail of solar-like ($T_{eff} \lesssim 6000$ K) multiplicity investigations, many studies have reported a paucity of confirmed low-mass stellar or 
brown dwarf companions with mass ratios ($q \equiv M_{c}/M_{*}$) $<$0.2 (see e.g. \citealt{pon05, bou11}), leading \citet{rag10} to suggest that short-period companions to solar-like stars prefer higher mass ratios, although there is disagreement in the literature regarding this trend (see e.g. \citealt{hal03}).
\citet{bur07} relate that the short-period brown dwarf desert seems to extend into the M dwarf regime.  \citet{rag10} also note that the majority of the companions surrounding solar-like stars with periods $<$100 days are triple systems, and thus potentially indicative that such systems experienced orbital migration \citep{bat02}.  Short-period low mass ratio binaries are much more commonly observed around slightly more massive F-type stars \citep{bou05,pon06,bea07,bou11}.  \citet{bou11} 
have suggested that low $q$ companions can form at or migrate to short orbital periods around a wide range of stellar primary masses, but suggest that they may not survive around G dwarfs and lesser mass primary stars. 

The Multi-object APO Radial Velocity Exoplanet Large-area Survey (MARVELS), one of the three surveys being executed during the Sloan Digital Sky Survey (SDSS) III \citep{eis11}, is a four-year program which is monitoring the radial velocities of $\sim$3,300 V=7.6-12 FGK type dwarfs and subgiants.  As described in \citet{lee11}, the target selection strategy attempts to impose minimal and well-understood biases on targets' ages and metallicities; hence, the 
survey provides an ideal, statistically robust means to explore the mass distribution of substellar and very low mass star companions over orbital periods of $\leq$2 years from a relatively low-biased target sample.  
In anticipation of a statistical analysis of global trends in the population of BDs and low mass binary companions identified by the survey, we are performing 
detailed and careful characterization of the fundamental parameters of these companions and their host stars (see e.g. \citealt{fle11} discussion of MARVELS-2b).  The first paper in this series was \citet{lee11}, which presented an analysis of a short period brown dwarf surrounding the F9 star TYC 1240-00945-1.  Analogous detailed characterization of individual 
systems has been shown to be particularly important in refining the radii of Kepler candidate planet-candidates to be more Earth-like than assumed \citep{mui11}.  The advantages of meta-analyzing such well-characterized systems is also demonstrated in \citet{bou11}, who were able to begin to explore the mass-radius relationship from 
the planetary to BD to very low mass star regimes as well as the mass ratio of companions as a function of primary mass.

In this paper, we present a detailed analysis of the fundamental properties of the solar-like star TYC 4110-01037-1 (hereafter TYC 4110) 
and report on the discovery of a very low mass stellar 
companion associated with the system.  In Section \ref{obs}, we describe the spectroscopic and photometric data which were used for this analysis.  We determine accurate 
fundamental stellar parameters for the star in Section 3 and describe the basic properties of the very low mass stellar companion in Section \ref{companion}.  Finally, 
we discuss the implications of these results in Section \ref{discuss}.

\section{Observations and Data Reduction} \label{obs}
\subsection{SDSS-III Radial Velocity Data}

Our primary radial velocity (RV) observations of TYC 4110 were obtained during the first two years of the 
SDSS-III MARVELS survey, which uses a dispersed fixed-delay interferometer \citep{ge09} on the SDSS 2.5m telescope \citep{gun06}.
A total of thiry-two observations were obtained over the course of $\sim$2 years.  Each 50 minute observation yielded
two fringing spectra (aka. ``beams'') from the interferometer spanning the wavelength regime $\sim$500-570 nm with R $\sim$12,000.  
\citet{lee11} describe the basic data reduction and analysis leading to the production of differential RVs for each beam of the 
interferometer.  We combined these beams for each observation set using a weighted mean.  As described in \citet{fle10}, we 
scaled up the RV errors by a ``quality factor'' (Q = 5.67 for TYC 4110) based on the rms errors of the other stars observed on the same SDSS-III plate
as TYC 4110.  

A summary of the relative amplitude RV measurements obtained for TYC 4110 with MARVELS is presented in Table \ref{rv}.

\subsection{TNG Follow-up Radial Velocity Data} \label{tng}

Supporting RV observations were obtained with the 3.6m Telescopio Nazionale Galileo (TNG) using its SARG spectrograph 
\citep{gra01}.  The 0$\farcs$8 x 5$\farcs$3 slit provided R $\sim$57,000 spectroscopy between 462-792 nm.  
We obtained six spectra with an iodine cell (IC), to provide high precision radial velocities (Table \ref{rv}), and one without the IC to serve as a stellar template. The data were reduced using standard IRAF routines, and RVs were measured using the IC technique (Marcy \& Butler 2000).  Each of 21 SARG spectral orders between 504-611 nm were divided in 10 pieces, and RV calculations were derived 
from each of the 210 resulting pieces.  Based on a goodness-of-fit indicator, the best 158 (75\%) pieces were selected. Following a 2-$\sigma$ clip, the remaining RV measurements were combined with a weighted 
average to produce the RV measurements quoted in Table \ref{rv}.

\subsection{APO Spectroscopic Data} \label{apo}

Two R$\sim$31,500 optical ($\sim$3,600 -10,000\AA) spectra of TYC 4110 were obtained on UT 2010 September 29 (HJD 2455468)
with the Apache Point Observatory 3.5m telescope and ARC Echelle Spectrograph (ARCES; \citealt{wan03}), to enable accurate characterization of stellar fundamental 
parameters.  The two spectra were obtained using the default 1$\farcs$6 x 3$\farcs$2 slit and an exposure time of 1200 seconds.  A ThAr lamp exposure was 
obtained between these integrations to facilitate accurate wavelength calibration.  The data were processed using standard IRAF techniques.  
Following heliocentric velocity corrections, each order was continuum normalized, and the resultant continuum normalized data from each observation was 
averaged.  The final spectrum yielded a signal-to-noise ratio (S/N) of $\sim$175 at $\sim$6500 \AA. 

\subsection{HAO Photometric Data}

We obtained absolute photometry of TYC 4110 using the Hereford Arizona Observatory (HAO), a private facility in Southern Arizona (observatory code G95 in the IAU Minor Planet Center).  HAO employs a 14-inch Meade LX200GPS telescope equipped with a SBIG ST-10XME CCD.  Observations in B, V, and I$_{c}$ filters 
were made on 2011 January 15 and 2011 February 10.  A total of 22 Landolt standard stars and 9 secondary standards based on Landolt star fields SA98 and SA114 \citep{lan92} were observed at several airmass values similar to the airmass for TYC 4110.  The target's magnitude was calculated by:
\begin{equation}
    M_{f} = M_{fo} - 2.5 \log_{10} (F_{f} / g) - (K_{f}^{'} \cdot m) + (S_{f} \cdot C)
\end{equation}
where M$_{f}$ is the magnitude for each filter $f$, $M_{fo}$ is a constant determined from the standard stars, $F_{f}$ is star flux using a large photometry aperture, $g$ is exposure time, $K_{f}^{'}$ is zenith extinction coefficient for each filter, $m$ is air mass, $S_{f}$ is the star color sensitivity (determined from the standard stars), and $C$ is star color (B-V).  Solutions for (B-V) were obtained by iterating V and B magnitudes from initial values 2-3 times.  The resultant absolute photometry for TYC 4110 is summarized in Table \ref{star}.

\subsection{SuperWASP Photometric Data}

We analyzed 5 epochs of broadband optical photometry (400-700 nm) of TYC 4110, obtained between 2006 April 7 and 2008 April 14, from the SuperWASP public 
archive \citep{but10}.  Aperture photometry of the 160 individual observations available in the archive, each taken with a 30 second integration time, was computed via the SuperWASP pipeline.  Further details about the observational design and data reduction pipeline for SuperWASP can be found in \citet{pol06}.  We find the SuperWASP photometry 
exhibits no statistically significant evidence of variability (error-weighted RMS $\sim$0.675\%) over the time-scales sampled by these data.  For example, a linear fit to the entire dataset yields negligible  
variation in flux with a best-fit slope of 0.040 $\pm$ 0.030\% day$^{-1}$.  We also detect no evidence of a transit.  We do caution however that the sampling of these data is sparse.  

\section{TYC 4110-01037-1: The Star} \label{star}
\subsection{Fundamental Stellar Properties} \label{stellarparam}

We analyzed moderate resolution spectroscopic data from the ARCES spectrograph using two separate analysis techniques to 
extract fundamental stellar parameters for TYC 4110.  We refer to these different pipeline results as the ``IAC'' (Instituto de Astrof\'{i}sica de Canarias) and ``BPG'' (Brazilian Participation Group) results, as described below.

\subsubsection{``IAC'' Analysis}

We derive equivalent widths (EWs) of \ion{Fe}{1} and \ion{Fe}{2} lines with the code {\scshape ARES} \citep{sou07}, using an initial linelist with 263 \ion{Fe}{1} and 36 \ion{Fe}{2} lines given in \citet{sou08}, and use the rules in \citet{sou08} to modify the \emph{rejt} parameter in {\scshape ARES} according to the signal-to-noise ratio of each spectrum.  In addition, we set the ARES parameters \emph{smoother} = 4, \emph{space} = 3 and \emph{miniline} = 2.  The parameter \emph{lineresol} was modified according to the resolving power of each spectrum.

The stellar atmospheric parameters were computed using the code {\scshape StePar} \citep{tab11}. This code employs the 2002 version of the MOOG code \citep{sne73}, and a grid of Kurucz ATLAS9 plane-parallel model atmospheres \citep{kur93}.  {\scshape StePar} iterates until the slopes of $A\left(\mathrm{Fe~I}\right)$ vs. $\chi$ and $A\left(\mathrm{Fe~I}\right)$ vs. $\log\left({\rm{EW}}~\lambda^{-1}\right)$ are equal to zero, while imposing the ionization equilibrium condition  $A\left(\mathrm{Fe~I}\right) = A\left(\mathrm{Fe~II}\right)$.  A 2-$\sigma$ rejection of the EWs of \ion{Fe}{1} and \ion{Fe}{2} lines is performed after a first determination of the stellar parameters, and then the {\scshape StePar} program is re-run without the rejected lines \citep[see][for further details]{tab11}.

For the ARCES spectrum of TYC 4110, 198 \ion{Fe}{1} lines and 24 \ion{Fe}{2} lines remain after clipping.  These are used to derive $T_{\rm eff} = 5879 \pm 25 $K, $\log{(g)} = 4.53 \pm 0.18$, $\rm{[Fe/H]} = -0.02 \pm 0.05$, and $v_{micro} = 0.932 \pm 0.038 \rm{km ~ s^{-1}}$.

Internal uncertainties were also derived for each stellar parameter.  The uncertainty of $v_{micro}$ was obtained by varying this parameter until the slope of the linear regression of $A\left(\mathrm{Fe~I}\right)$ versus $\log\left({\rm{EW}}~\lambda^{-1}\right)$ was equal to its standard deviation.  The uncertainty of $T_{\rm eff}$ was determined by changing this parameter until the slope of the linear regression of $A\left(\mathrm{Fe~I}\right)$ versus $\chi$ was equal to its standard deviation.  The uncertainty of $v_{micro}$ was also taken into account when calculating the uncertainty of $T_{\rm eff}$.  The uncertainty of $\log{(g)}$ was obtained by varying this parameter until the difference between the mean abundances from \ion{Fe}{1} and \ion{Fe}{2} were equal to the standard deviation of the latter.  The contributions from $T_{\rm eff}$ and $v_{micro}$ were included.  Finally, the uncertainty of $\rm{[Fe/H]}$ is a combination of the standard deviation of the \ion{Fe}{1} abundance and the variations caused by the errors in $T_{\rm eff}$, $\log{(g)}$ and ${\xi}_{t}$, all added in quadrature.

\subsubsection{``BPG'' Analysis}
We assume LTE and use the 2002 version of MOOG \citep{sne73}, along with the 1D plane-parallel model atmospheres interpolated from the ODFNEW grid of ATLAS9 models \citep{cas04}.  Initially, a list of $\sim$150 isolated and moderately strong (i.e., $5 < \rm{EW} < 120$ mA) \ion{Fe}{1} and \ion{Fe}{2} lines was compiled using the Solar Flux Atlas \citep{kur84}, the Utrecht spectral line compilation \citep{mo66}, and a Ganymede ARCES spectrum with $\rm{S/N} = 400$. The values for the central wavelengths and line excitation potentials were taken from the Vienna Atomic Line Database (VALD; \citealt{kup99}). We also multiplied the van der Waals damping parameter ``C6'' by a factor of two, following \citet{hol91}.

The EWs of these lines were automatically measured in the solar spectrum from fits of Gaussian profiles using the task \emph{bplot} in IRAF. The quality of the measurements was checked by performing two tests.  First, since the line depth is expected to be a linear function of the reduced EW ($\rm{EW} \lambda^{-1}$) for non-saturated lines, we 
eliminated lines that did not follow a linear relation, using a 2-$\sigma$ clipping.  The second test is based on the fact that the shapes of the lines are essentially determined by the instrumental profile at the APO resolution ($\sim$31,500). Since the resolution is approximately constant over the entire spectrum, we expect the quantity $\rm{FWHM} \lambda^{-1}$ to be approximately constant for lines of the same species.  We therefore perform a linear fit to this relation and eliminate lines that exhibit 2-$\sigma$ deviations.

After these tests, the final solar line list contained 91 \ion{Fe}{1} and 11 \ion{Fe}{2} lines. Solar $gf$ values were derived for all these lines using a solar model atmosphere with the following parameters: $T_{\rm eff} = 5777$ K, $\log{(g)} = 4.44$, $\rm{[Fe/H]} = 0.0$ and $v_{micro} = 1.0 ~ \rm{km ~ s^{-1}}$. The adopted solar abundance for iron is A(Fe) = 7.50 \citep{asp09}.

EWs for 102 Fe lines were measured in the TYC 4110 APO spectrum and checked with the tests described above.  $T_{\rm eff}$ and $v_{micro}$ were iterated until zero slopes were found in the plots of A(\ion{Fe}{1}) versus $\chi$ and $\log{\left(\rm{EW} ~ \lambda^{-1}\right)}$, respectively; i.e., until the individual \ion{Fe}{1} line abundances were independent of excitation potential and reduced EWs. The surface gravity was iterated until A(\ion{Fe}{1}) = A(\ion{Fe}{2}), i.e., until the same average abundances were given by \ion{Fe}{1} and \ion{Fe}{2} lines.  At the end of this iterative process, a consistent set of atmospheric parameters ($T_{\rm eff}$, $\log{(g)}$, $[Fe/H]$, and  $v_{micro}$) was obtained for the star.  Note that the metallicity is simply given by [Fe/H] = $A\left(\mathrm{Fe}\right) - 7.50$, where 7.50 is the solar iron abundance taken from \citet{asp09}. At this point, any lines with abundances that deviated more than 2-$\sigma$ from the average were removed and the above iteration was repeated until convergence was achieved. 

We derive $T_{\rm eff} = 5878 \pm 49 $ K, $\log{(g)} = 4.43 \pm 0.17$, $\rm{[Fe/H]} = 0.00 \pm 0.06$ and $v_{micro} = 1.00 \pm 0.08 ~ \rm{km ~ s^{-1}}$ based on the ARCES spectrum.  The final line list after rejections contained 60 \ion{Fe}{1} and 8 \ion{Fe}{2} lines.  The internal uncertainties are calculated in the same was as the ``IAC'' analysis above.

\subsubsection{Final Stellar Parameters}
We determined the mean values for $T_{\rm eff}$, $\log{(g)}$, $[Fe/H]$, and $v_{micro}$ by combining the results from the IAC and BPG analyses via a mean, weighted by the inverse of the internal variances.  For each parameter, we add in quadrature a systematic error of 18 K, 0.08, 0.03 and 0.02 $\rm{km~s^{-1}}$ for $T_{\rm eff}$, $\log{(g)}$, $[Fe/H]$, and $v_{micro}$, respectively.  These systematic errors are calculated based on the weighted standard deviation of the weighted means of each parameter using 18 spectra of 13 stars (7 MARVELS targets and 6 stars with well-known atmospheric parameters).  These stars span $T_{\rm eff}$ from 5200-6500 K, $\log{(g)}$ from 4.0-4.7, $[Fe/H]$ from -0.5 - +0.5 and $v_{micro}$ from 0.3 - 1.8 $\rm{km~s^{-1}}$.  The final stellar parameters are T$_{eff}$ = 5879 $\pm$29 K,  \hbox{$\log g$} = 4.48 $\pm$0.15,  $[Fe/H]$ = -0.01 $\pm$0.05, and $v_{micro}$ = 1.40 $\pm$0.05 km s$^{-1}$ (Table \ref{star}).  

\begin{figure}
\begin{center}
\includegraphics[width=12cm]{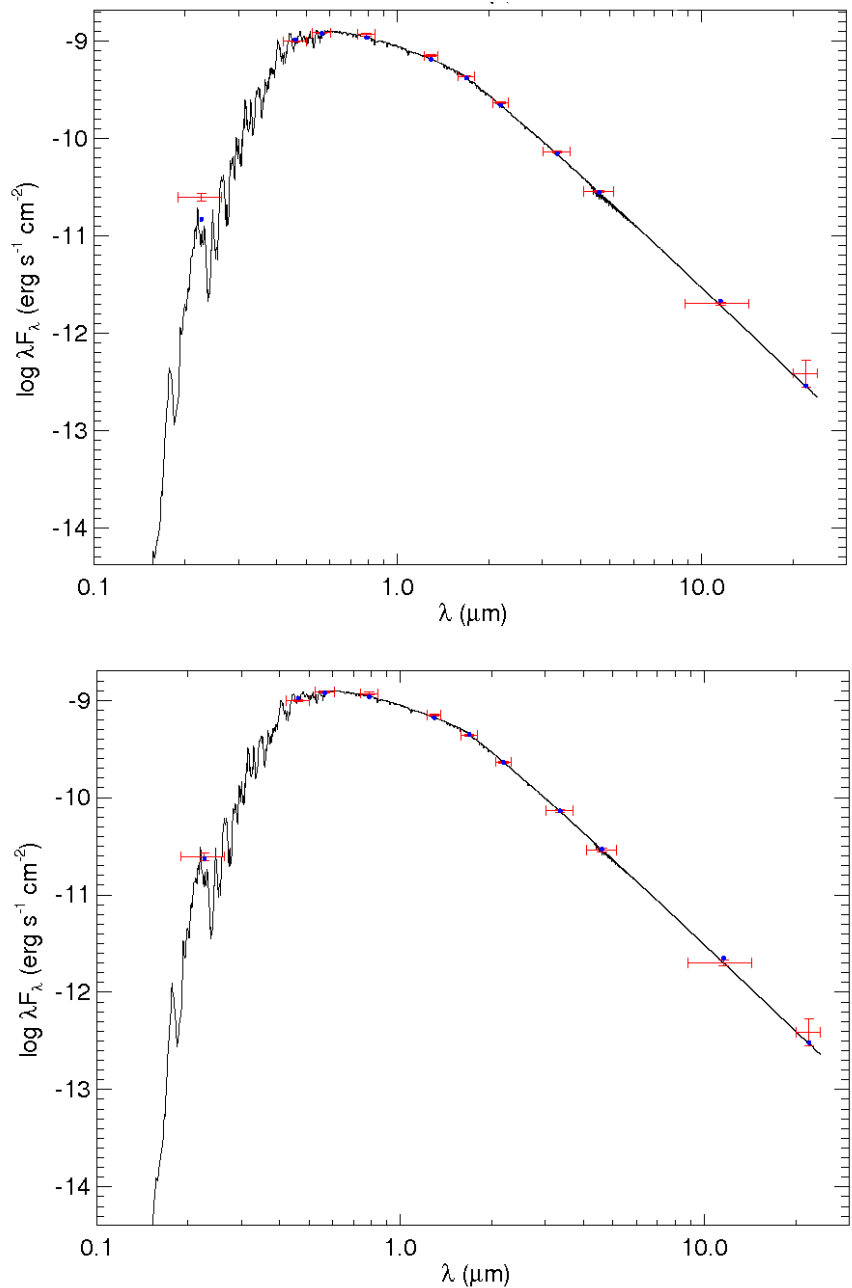}
\caption{\textit{Top:} The observed near UV through IR SED for TYC 4110-01037-1 is shown along with a best-fit NextGen model atmosphere.  The resultant 
fundamental stellar parameters from this fit agreed to within 1-$\sigma$ with the stellar parameters determined from analysis of moderate resolution spectra (Section \ref{stellarparam})  \textit{Bottom:} By constraining T$_{eff}$ and log (g) to the spectroscopically determined values during the SED fit, we are able to better constrain the 
total line-of-sight extinction to be A$_{V}$ = 0.16 $\pm$0.04.  Using this A$_{V}$, we estimate the distance to TYC 4110-01037-1 to be 125.1 $\pm$4.6 pc.
 \label{sed}}
\end{center}
\end{figure}

To check these parameters using a different observational technique, we constructed a spectral energy distribution (SED) for TYC 4110, using the near UV to 2MASS \citep{cut03} and WISE \citep{wri10} IR photometry compiled in Table \ref{star} (see Figure \ref{sed}).  These data were fit with a NextGen model atmosphere \citep{hau99}, and we limited the maximum line-of-sight extinction to be $A_{V}$ to $<$ 0.28 from analysis of dust maps from \citet{sch98}.  The resultant 
parameters, T$_{eff}$ = 6000 $\pm$200 K, \hbox{$\log g$} = 4.0 $\pm$1.0, $[Fe/H]$ = 0.0 $\pm$0.3, and $A_{V}$ = 0.20$\pm$0.08, all agree to within 1-$\sigma$ of the results found via analysis of our moderate resolution spectroscopy.  We also fit these SED data by constraining the T$_{eff}$ and \hbox{$\log g$} values to those derived from our spectroscopic analysis.  The resultant fit (reduced $\chi^{2}$ = 2.62; see panel b Figure \ref{sed}) provides a more robust estimate of $A_{V}$, 0.16 $\pm$0.04.  Using this total extinction estimate, and adopting a $BC_{V}$ of -0.19  $\pm$0.02 \citep{cox00}, we estimate the distance to TYC 4110 to be 125.1 $\pm$4.6 pc (Table \ref{star}).

\subsection{Stellar Mass and Radius} \label{mass}

Using the spectroscopically determined values of T$_{eff}$, \hbox{$\log g$}, and $[Fe/H]$ (Table \ref{star}), we determined the mass and radius of TYC 4110 using the 
empirical \citet{tor10} relationship.  We find M$_{\star}$ = 1.07 $\pm$0.08 M$_{\Sun}$ and R$_{\star}$ = 0.99 $\pm$0.18 R$_{\Sun}$ (Table \ref{star}).  Our quoted errors include contributions from the uncertainties in our fundamental stellar parameters, as well as the scatter in the \citet{tor10} relationship ($\sigma_{logm}$ = 0.027 and $\sigma_{logr}$ = 0.014) and correlations of the best-fit coefficients from \citet{tor10} added in quadrature.  We did not include covariances between T$_{eff}$, $\log g$, and $[Fe/H]$ in this error analysis; however, our final quoted uncertainties for these values do include a systematic error term that conservatively encapsulates any 
covariance between these parameters.  We did however perform a Monte 
Carlo simulation of our spectroscopically determined stellar parameters and the \citet{tor10} relations, and found a stellar mass and radius consistent with the 
aforementioned values.

\subsection{Evolutionary State} \label{evolve}

\begin{figure}
\begin{center}
\includegraphics[width=12cm]{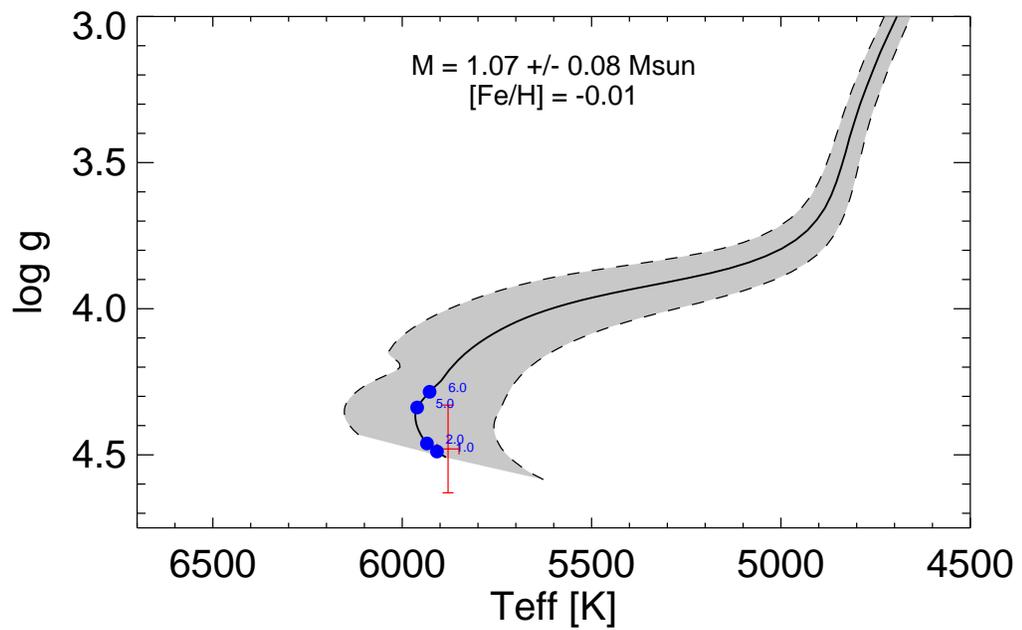}
\caption{The observed stellar parameters for TYC 4110-01037-1 (red data) are compared to a Yonsei-Yale stellar \citep{dem04} 
evolutionary track for a M$_{\star}$ = 1.07 M$_{\Sun}$ star with $[Fe/H]$ = -0.01.  
Ages of 1.0, 2.0, 5.0, and 6.0 are indicated in blue, and 1-$\sigma$ deviations in the evolutionary track are shown in the shaded 
region. \label{hrd}}
\end{center}
\end{figure}

We assess the evolutionary state of TYC 4110 by comparing its spectroscopically measured fundamental stellar parameters against a Yonsei-Yale stellar 
evolutionary track \citep{dem04} for a M$_{\star}$ = 1.07 M$_{\Sun}$ star having $[Fe/H]$ = -0.01.  This is done in Figure \ref{hrd}, where the 
shaded region depicts deviations in the evolutionary track which 
would be expected from a 1-$\sigma$ (0.08 M$_{\Sun}$) change in the assumed stellar mass, while circles denote different time stamps in the track.  TYC 4110 lies near a predicted age of $\lesssim$5 Gyr in Figure \ref{hrd}; we therefore conclude that it is a main sequence dwarf star.  The lack of any detectable Ca II H and K emission in our ARCES spectra ( log(R$_{HK^{'}}$) $\sim$ -5.1) indicates the star is relatively inactive, and thus qualitatively consistent with the evolutionary state we derive.

\subsection{Systemic and Rotational Velocity}

We computed the absolute radial velocity of TYC 4110 by cross-correlating the six epochs of SARG spectra against a Solar spectrum.  After removing the radial velocity contribution at each epoch induced by the presence of TYC 4110's companion (Table \ref{rv}), we determine the systemic velocity of TYC 4110, v$_{systemic}$, to be 
25.9 $\pm$0.2 km s$^{-1}$ (Table \ref{star}).  We also compared these SARG data to broadened versions of Kurucz ATLAS synthetic spectra to constrain the rotational velocity, 
v$_{rot} sin i$, of TYC 4110.  After considering a range of macroturbulence values, we find v$_{rot} sin i$ $\lesssim$3 km s$^{-1}$, which is slightly below the level of 
instrumental broadening present in these data ($\sim$5.3 km s$^{-1}$).

\section{TYC 4110-01037-1's Companion} \label{companion}
\subsection{Binary Companion Detection}

Analysis of post-pipeline processed MARVELS data involves searching for periodic behavior which is consistent with Keplerian orbital motion, and filtering out 
``contaminant'' RV signals which do not arise from the presence of a companion.  One of the first steps in this process is to compute a Lomb-Scargle (LS)
periodogram \citep{lom76,sca82} and search for periodic signals which significantly exceed a conservative false alarm probability assessment.  There have been several implementations of this method \citep{pre89}; we followed \citet{cum04} which uses the comparison of the $\chi^{2}$ values between a sinusoidal fit at a given frequency and a fit to the mean to generate the power term (see e.g. Cumming's equation 2).  
This form was chosen to be fully extensible by adding linear terms, harmonic terms, or even a fully Keplerian fit.  We 
stepped through frequency space using steps and a search window as 
appropriate for our data sampling, as described in \citet{pre92}.  In order to interpret the significance of this power spectrum, false alarm probabilities 
(FAP) were calculated using the techniques in \citet{bal08}.  

Data from the current MARVELS pipeline have systematic effects present which can mimic a companion signal.  We can significantly mitigate this issue by taking
advantage of the fact that the MARVELS instrument observes 60 stars at a time.  As a result, we can search for periodic signals from all stars on a given plate, 
and use any detected signals to characterize systematics in our data.  This is done by taking the sum of the power for each frequency across the entire plate.  We then remove the highest power from each frequency (so that actual companions do not skew the average), excluding any power associated with the candidate companion, 
and compute the average.  

\begin{figure}
\begin{center}
\includegraphics[width=14cm]{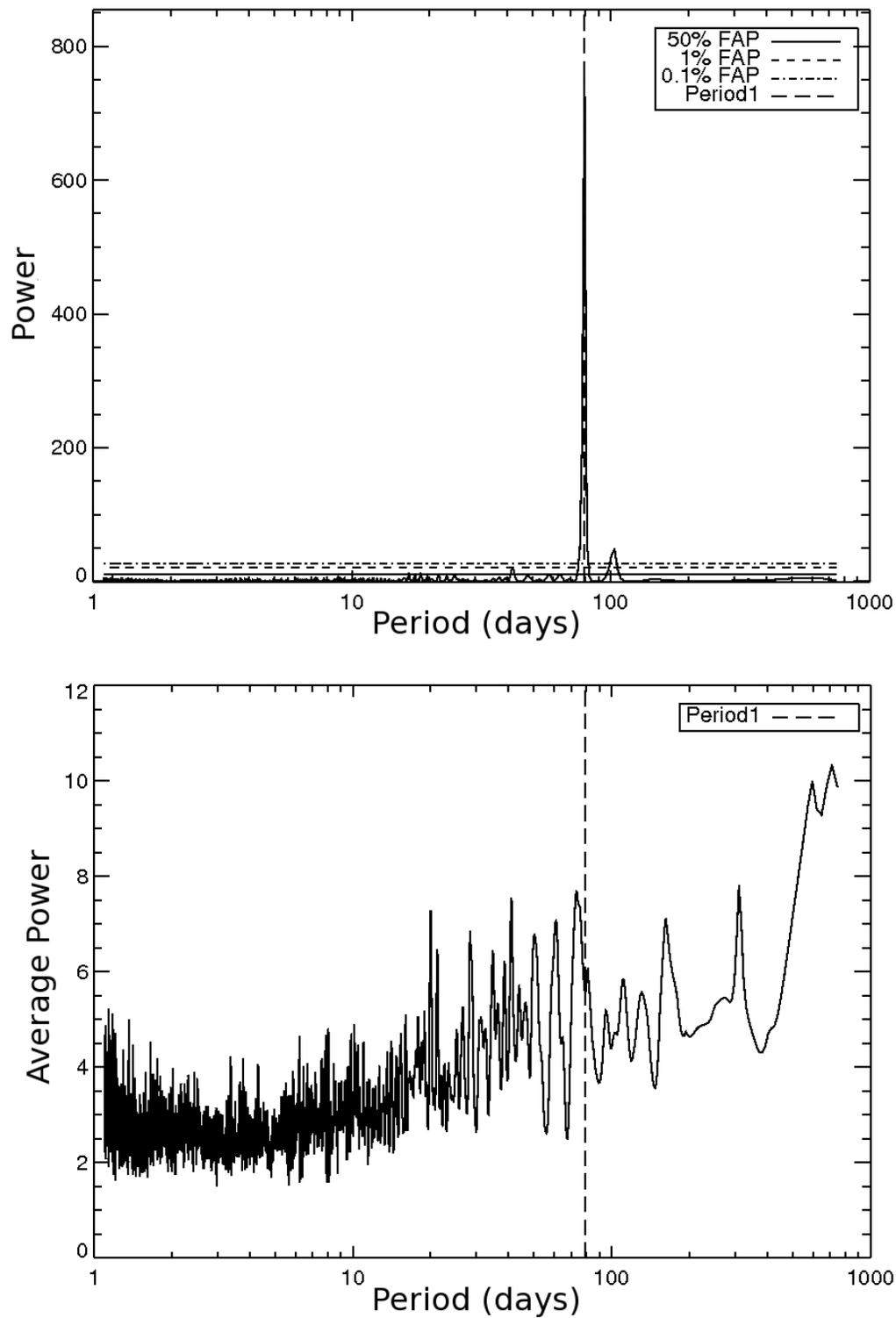}
\caption{The periodogram for RV measurements of TYC 4110-01037-1 (panel a) indicates the strong presence of a $\sim$79-day periodic signature, indicated by the dashed vertical line.  
This signal is not seen to coincide with any strong feature in the 
periodogram for all stars located on the same SDSS-III plate as this object (panel b), indicating it is not caused by a known artifact of our instrument or reduction 
pipeline.  \label{perio}}
\end{center}
\end{figure}

Figure \ref{perio} indicates TYC 4110 exhibits a strong $\sim$79-day period signal.  The companion to TYC 4110, hereafter referred to as 
MARVELS-3B, does not match any strong periodic signature identified in the power spectrum of all stars located on the same SDSS-III plate (bottom panel; Figure \ref{perio}), and is 
therefore not caused by any known systematic artifact in our instrument or reduction pipeline.  

\subsection{Radial Velocity Fits}

Radial velocities derived from MARVELS and SARG data were fit with the EXOFAST code \citep{eas11} to extract detailed Keplerian orbital parameters.  We first performed an independent fit of the MARVELS data, and re-scaled the MARVELS error bars to force the probability of $\chi^{2}$, P($\chi^{2}$), = 0.5.  Since we did not have enough SARG data points to perform an independent fit solely on these data, we then fit the combined (MARVELS + SARG) data, and re-scaled the SARG errors to force P($\chi^{2}$) = 0.5.  The fitting of the combined dataset and re-scaling of the SARG errors were iterated until a convergent solution was achieved.  The MARVELS and (MARVELS + SARG) data yielded consistent results to within 1-$\sigma$ of the fit errors.  We also computed the $\chi^{2}$ of the SARG data about the MARVELS-only fit, and the resultant value ($\chi^{2}$ = 8.47 for 5 degrees of freedom) indicates such a fit would only happen by chance $\sim$13\% of the time.  We hereafter only consider the combined MARVELS + SARG fit.  

The raw MARVELS and SARG radial velocities were computed on independent, relative scales.  We determined that the best offsets to simultaneously analyze these data about a zero-point of 0 m s$^{-1}$ were 1345$\pm24$ m s$^{-1}$ (SARG) and 8583 $\pm$12 m s$^{-1}$ (MARVELS); the RVs quoted in Table \ref{rv} have had these offsets applied.  The RV errors listed in Table \ref{rv} include the re-scaling factors described above.  The inclusion of an additional linear term was explored in the fitting process, but there is no compelling evidence for an acceleration due to another companion, with a best-fit linear slope of 0.096$_{-0.040}^{+0.039}$ m s$^{-1}$ day$^{-1}$ observed, corresponding to a 2.4-$\sigma$ deviation. 

\begin{figure}
\begin{center}
\includegraphics[width=14cm]{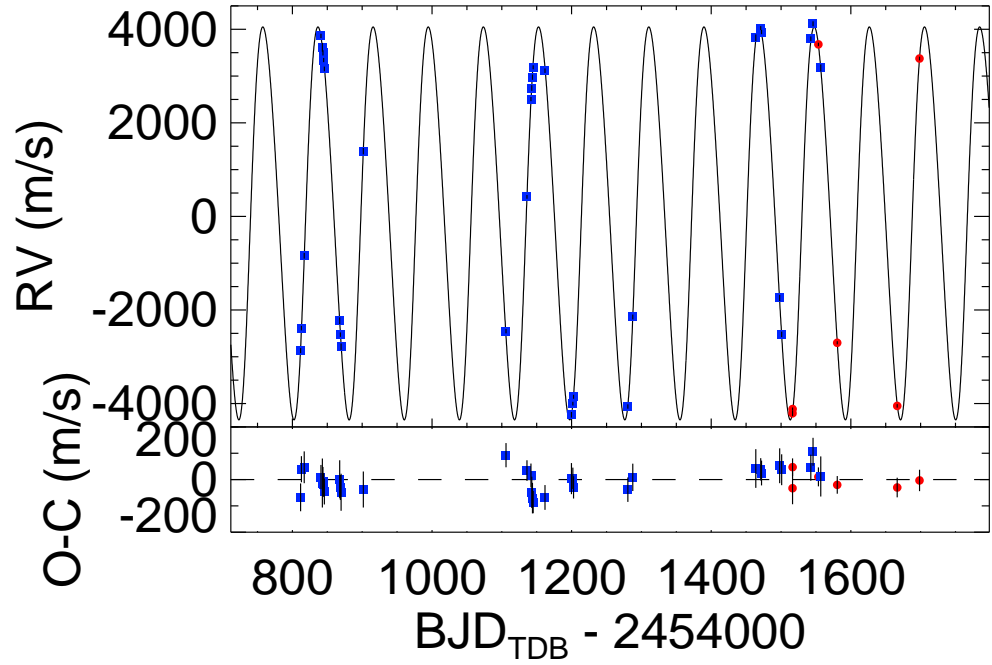}
\caption{The derived relative radial velocities from the MARVELS (blue points) and SARG (red points) spectrographs are overlayed with the best fit 
orbital solution described in Section \ref{companion} and compiled in Table \ref{fit}.  Residuals to this fit are shown in the bottom panel.  \label{rvs}}
\end{center}
\end{figure}

\begin{figure}
\begin{center}
\includegraphics[width=14cm]{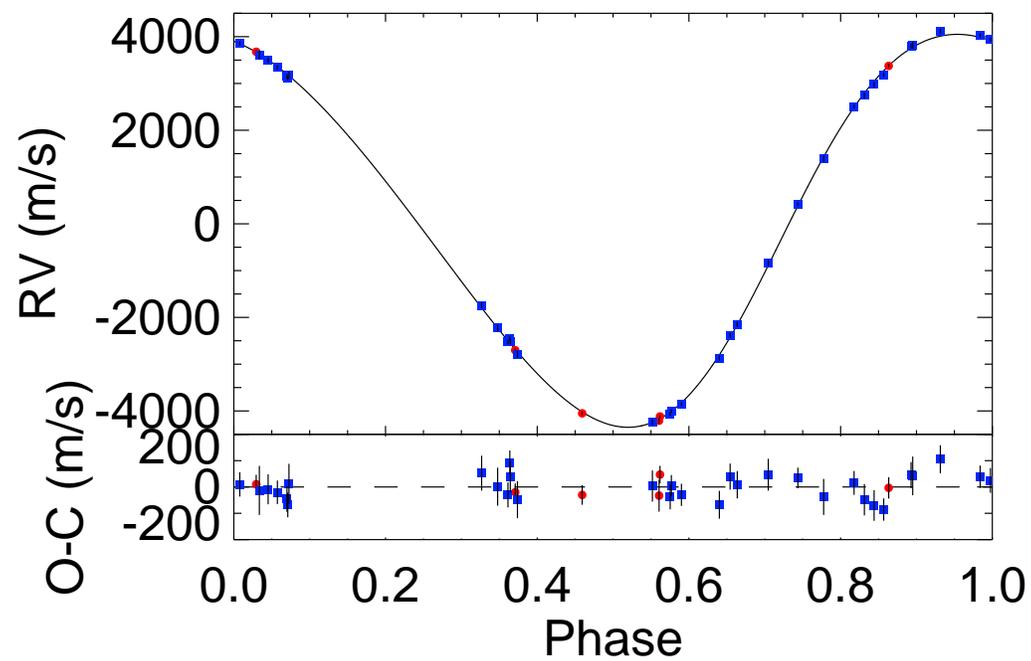}
\caption{The phase-folded radial velocity curve is shown, for a MARVELS-3B period of  78.994$\pm$0.012 days and eccentricity of 0.1095 $\pm$0.0023.  Based 
on the derived stellar mass of 1.07 $^{+0.08}_{-0.08}$ M$_{\Sun}$ (Table \ref{star}), we determine the minimum ($\sin i$ = 1) mass of MARVELS-3B to be 
$>$ 97.7$\pm$5.8 M$_{Jup}$. \label{phased}}
\end{center}
\end{figure}

As seen in the full RV curve (Figure \ref{rvs}) and in the phase-folded curve (Figure \ref{phased}), these data are described by a 78.994 $\pm$0.012 day period, a moderately elliptical ($e$ = 0.1095 $\pm$0.0023) orbit, and a semi-amplitude $K$ = 4199 $\pm11$.  Full fit parameters for TYC 4110 are presented in Table \ref{fit}.  To search for 
evidence that MARVELS-3B was a transiting system, we computed a LS periodogram of the available SuperWASP photometry, but found no evidence of variability at the $\sim$79 day period of MARVELS-3B above a level of $\sim$0.8\%.

We determined the mass for MARVELS-3B using:
\begin{equation}
\frac{(M_{c}\sin i)^{3}} {(M_{*}+M_{c})^{2}} = \frac{K^{3}  (1-e^{2})^{(3/2)}  P} {2 \pi  G}
\end{equation}
\begin{equation}
\frac{(M_{c}\sin i)^{3}} {(M_{*}+M_{c})^{2}}  = (5.953 \pm 0.047) \cdot 10^{-4} M_{\Sun} 
\end{equation}

\noindent and the fit parameters compiled in Table \ref{fit}.  This yields a minimum companion mass (if $\sin i$ = 1)  of 97.7 $\pm$5.8 M$_{Jup}$, which places MARVELS-3B slightly above the generally accepted brown dwarf upper mass limit of 80 M$_{Jup}$ and into the low-mass star regime.  The minimum mass ratio, $q$, of the companion to the primary is 0.087 $\pm$0.003.

\subsection{Binary Companion Mass}

The true mass of the companion depends on the inclination $i$ of the orbit, which is unknown.  However, we can estimate the posterior probability distribution of the true companion mass, given an isotropic distribution of orbits, and adopting a prior for the distribution of the companion mass ratios.  We proceed to do this using a Monte Carlo method, following the methodology described in detail in \citet{fle10} and \citet{lee11}, which we briefly summarize here.  We combine the posterior distribution of orbital parameters $K$, $e$, and $P$ obtained from the Markov Chain Monte Carlo (MCMC) fit to the radial velocity data, with an estimate of the joint distribution of the primary mass and radius obtained using the spectroscopically-determined $T_{eff}$, \hbox{$log g$} and $[Fe/H]$ combined with the \citet{tor10} relations, accounting for all sources of uncertainty in the measured values and the relations themselves.  We draw values of $\cos{i}$ from a uniform distribution.  The values of $K$, $e$, and $P$ determine the mass function $(M_{c}\sin{i})^{3}/(M_{*}+M_{c})^{2}$, and then the value of $i$ along with the primary mass $M_{*}$ determines $M_{c}$. Finally, we appropriately weight the resulting distribution of $M_{c}$ by our prior on the mass ratio $q$.  

As described in Section \ref{intro}, the mass ratio of companions 
around G dwarfs is not well constrained by current observations and MARVELS-3B likely lies in a relatively underpopulated region of mass ratio parameter space.  Nevertheless, 
we consider several different priors on the companion mass ratio which we suggest are reasonable given current observations (see e.g. \citealt{gre06}), of the form:
$dN/dq$ $\propto$ q$^{+1}$, $dN/dq$ $\propto$ q$^{-1}$, and $dN/dq$ = constant.  We note that massive companions are ruled out by the lack of a statistically significant infrared excess in the SED (Figure \ref{sed}) and the lack of a secondary component in our high-resolution optical spectra (see discussion below).  We include this constraint by weighting the resulting distribution of $M_{c}$ by $exp[-0.5(\Delta K/\Delta K_{max})^{2}]$, where $\Delta K_{max}$ is the upper limit on the excess flux (in magnitudes) in the $K$ band, and $\Delta K$ is the excess flux contributed by a companion of mass $M_{c}$ and a primary of mass $M_{*}$, as determined using the \citet{bar98} solar-metallicity, $Y=0.275$, 1 Gyr mass-magnitude relations (note that we could have adopted any isochrone in the range of 1-10 Gyr with negligible difference).  We caution the reader that this specific constraint does not fully and uniformly represent the prior for every possible 
configuration of the companion, but rather is a reasonable, simplying set of conditions.  We did not observe a statistically significant IR excess flux in TYC 4110's SED (Figure \ref{sed}); thus, we used the 3-$\sigma$ standard deviation (0.06 magnitudes) of the Ks-band data for $\Delta K_{max}$.  We note that the IR flux contribution from the wider separation candidate tertiary companion, discussed in more detail in Section \ref{tertiary}, is less than this $\Delta K_{max}$ and therefore does not influence our mass estimate.

\begin{figure}
\begin{center}
\includegraphics[width=14cm]{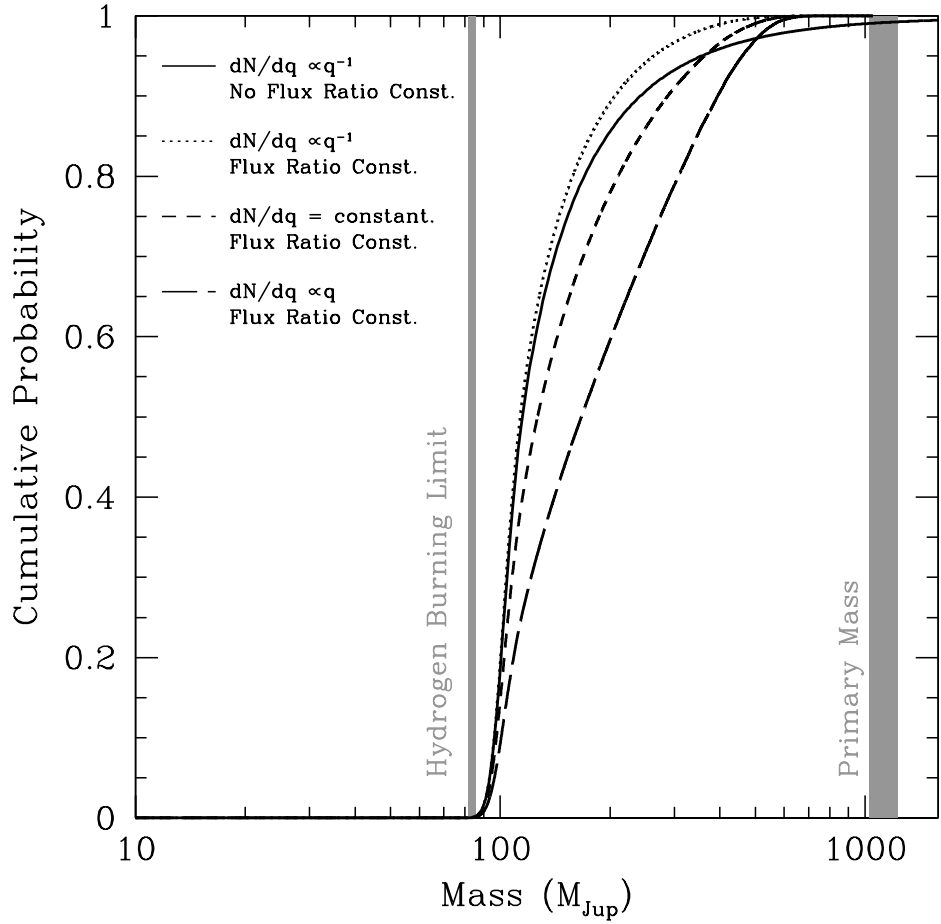}
\caption{The cumulative probability that the mass of MARVELS-3B is less than a given mass is shown, for 4 different priors on the companion mass ratio described in 
Section \ref{companion}.  \label{mcmc}}
\end{center}
\end{figure}

The resultant cumulative distributions of the true mass are shown in Figure \ref{mcmc}, and we summarize the median mass for each of these priors, along with 
the resultant mass ratio, in Table \ref{tableofq}.  Since these probabilistic median values rely on the simplifying assumptions we have made, we have not assigned formal 
confidence limits to these values, to prevent over-interpretation of their robustness.  For each prior, we note that the 
resultant median mass is in the M dwarf region with a low ($<$0.2) mass ratio, and companions more massive than 0.5 M$_{\sun}$ are ruled out at the 95\% probability level.

\begin{figure}
\begin{center}
\includegraphics[width=14cm]{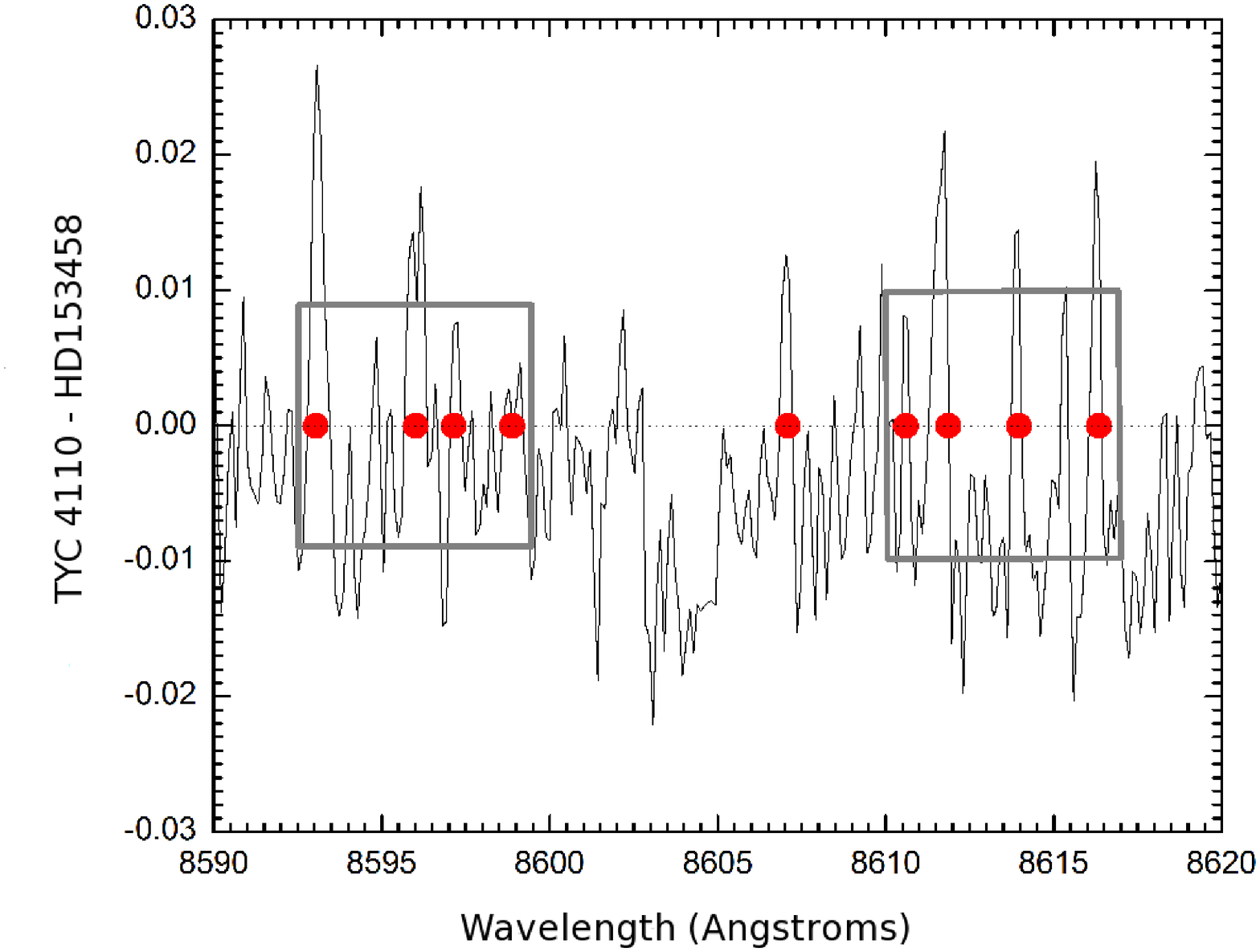}
\caption{The continuum normalized spectrum of HD153458 was subtracted from the continuum normalized spectrum of TYC 4110-01037-1.  The 1-$\sigma$ Poissan errors in the data are illustrated with grey boxes.   As discussed in Section 4.3, we note that minor mismatches between the properties of the reference and science spectra could produce noticable subtraction residuals in line wings, which suggests that only strong, repeatable deviations in these difference spectra should be interpreted as real contributions from companion(s).  A M3-type companion would have contributed a 2\% flux
enhancement to the system.  However, the difference spectrum does not exhibit $>$3-$\sigma$ deviations above the level of the Poisson noise (0.8\%) and continuum normalization uncertainties present in the data, which sets the lower mass limit for MARVELS-3B that can be ascertained from these specific data. \label{gus}}
\end{center}
\end{figure}

Finally, we also analyzed the ARCES spectrum of TYC 4110, obtained at a phase of $\sim$0.71, to search for evidence of MARVELS-3B.  We first note that we detected no evidence that the system was a SB2, which suggests $q$ $\lesssim$ 0.65 \citep{hal03}, i.e. MARVELS-3B is less massive than a mid K-type dwarf.  
Figure \ref{gus} illustrates the red optical difference spectrum computed by subtracting the spectrum of a G dwarf with very similar fundamental stellar parameters, HD 153458, from TYC 4110.  Both the known binary companion to TYC 4110 (MARVELS-3B) and the candidate tertiary companion (see Section \ref{tertiary}) could fill in the spectral features identified with red dots in Figure \ref{gus}, leading to positive deviations in the difference spectrum, although differentiating the precise amount of any contribution from the secondary versus the candidate tertiary is not possible.  While calibrating our usage of this technique, we found that minor mismatches between the properties of the reference and science spectra could produce noticable subtraction residuals in line wings (as seen in Figure \ref{gus}), which suggests that only strong, repeatable deviations in these difference spectra should be interpreted as real contributions from companion(s).  Using spectra from the \citet{pic98} library, we estimate that a M3-type companion would have contributed a $\sim$2\% flux enhancement to the system.  The difference spectrum does not exhibit $>$3-$\sigma$ deviations above the level of the Poisson noise (0.8\%) and continuum normalization uncertainties present in the data, which sets the lower mass limit for MARVELS-3B that can be ascertained from these specific data.

\subsection{Candidate Tertiary Companion} \label{tertiary}

\begin{figure}
\begin{center}
\includegraphics[width=14cm]{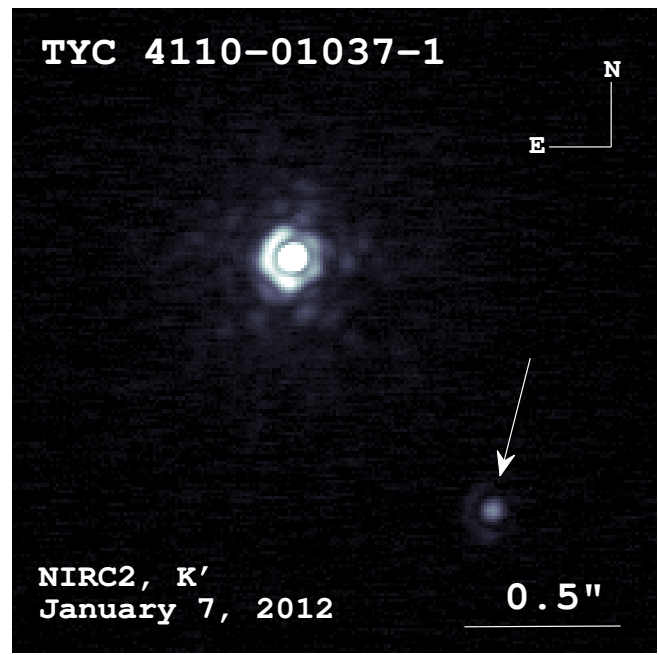}
\caption{Keck adaptive optics image of TYC 4110-01037-1 in K$^{'}$.  We have detected a faint candidate tertiary companion (indicated by the arrow) with red colors that is
separated by $986+/-4$ mas from the primary star.  If it is physically associated with the primary, it is most likely a dM3-dM4 star. \label{ao}}
\end{center}
\end{figure}

Triple star systems are a relatively common outcome of the star formation process (see e.g., \citealt{tok04}).  To further assess the multiplicity of TYC 4110, we acquired
adaptive optics (AO) images on 2012 January 7 using NIRC2 (PI: Keith Matthews) at the Keck Observatory in natural guide star mode. Our initial data set consisted of nine dithered images taken in the K$^{'}$ filter.  Inspection of the raw frames showed evidence for a faint candidate
companion located to the south-west of TYC 4110.  Figure \ref{ao} shows the fully processed K$^{'}$' image. 

We measured an accurate position for the candidate companion using the technique described in \citet{cre12}.  We first fit Gaussian functions 
to the stellar and companion point-spread-functions to locate their centroids in each frame.  The primary star was not saturated
in any of our dithered images.  We then correct for distortion in the NIRC2 focal plane (narrow camera mode) using the publicly available software
provided by the Keck Observatory astrometry support page \footnote{$http://www2.keck.hawaii.edu/inst/nirc2/forReDoc/post$\textunderscore$observing/dewarp/$}.
The results are averaged and the uncertainty in the separation and position angle is taken as the standard deviation, taking into account
uncertainty in the plate scale and orientation of the array by propagating these errors to the final calculated position.  Adopting a plate 
scale of $9.963 \pm 0.006$ mas pixel$^{-1}$ and instrument orientation relative to the sky of $0.13^{\circ} \pm 0.02^{\circ}$, as 
measured by \citet{ghe08}, we find a companion separation and position angle of $\rho=986\pm4$ mas and $PA=218.1^{\circ} \pm 0.3^{\circ}$
respectively.   

Upon noticing the companion, we obtained additional images in the J and H-bands to facilitate characterization.  Our aperture photometry indicates that the object
has red colors: $\Delta J = 4.219 \pm 0.104$, $\Delta H = 3.940 \pm 0.032$, and $\Delta K' = 3.805 \pm 0.027$ mags. Table \ref{3table} lists its apparent magnitude as measured relative to the primary star, taking into account the combined light from each source.

Both the (J-H) = 0.55 and (J-K$^{'}$) = 0.75 colors indicate a spectral type of $\sim$M3V \citep{leg02}.  Assuming the candidate is situated at the same distance as the primary, we find that the absolute magnitudes, $M_J=8.10 \pm 0.21$, $M_H=7.55 \pm 0.32$,
$M_K=7.36 \pm 0.13$, are each consistent with a $\sim0.25M_{\odot}$ star when
compared to the \citet{gir02} evolutionary models, which corresponds to a main-sequence spectral type of $\sim$dM4.  The relations from Table 5 of
\citet{kra07} yield the same result (dM4), though with a possibly lower mass estimate of $\sim 0.20M_{\odot}$.   

Given the RA and DEC and distance of TYC 4110, the a priori likelihood of detecting a background star within 1$\farcs$0 is $\sim$0.8\%.  With a proper motion of [-0.20, -99.40] mas yr$^{-1}$, a time baseline of several months will be sufficient to assess its association with the primary star.  We note that the detection and mass constraints placed on MARVELS-3B in this paper hold regardless of whether or not this candidate tertiary companion is confirmed to be co-moving with the primary star.

\section{Discussion} \label{discuss}

We discuss some of the derived properties for MARVELS-3B in the context of previous studies of low-mass companions to solar-like stars.  Both \citet{duq91} and \citet{rag10} demonstrate that companions to solar-like stars having orbital periods $\leq$12 days are circular; however, companions having orbital periods similar to that of MARVELS-3B, $\sim$79 days, exhibit eccentricities from 0-0.6 \citep{duq91,may92,rag10}.  MARVELS-3B's modest eccentricity of $\sim$0.11 therefore 
is clearly consistent with that observed for other similar period companions.

\begin{figure}
\begin{center}
\includegraphics[width=10cm]{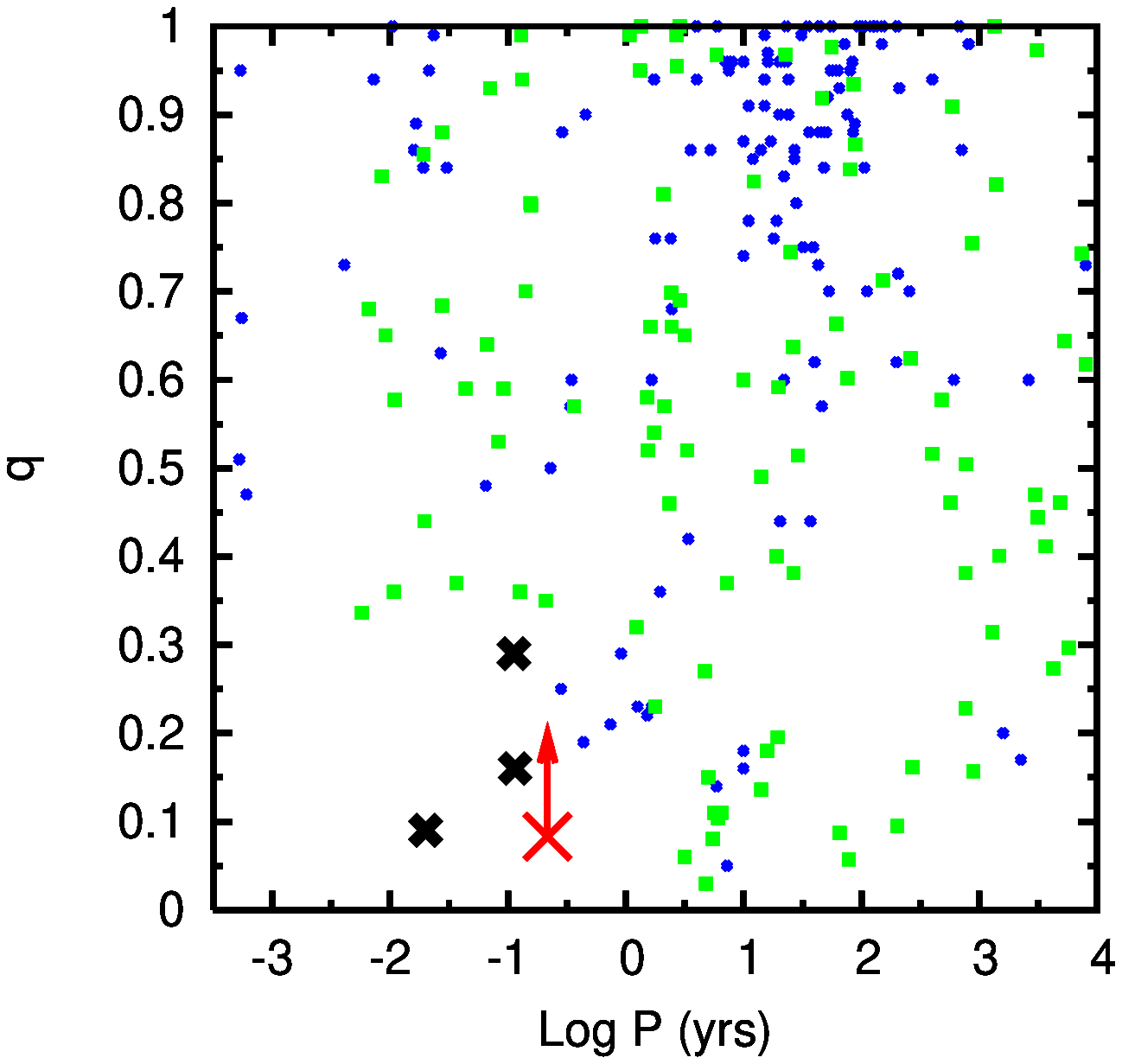}
\caption{The mass ratio, $q$, versus orbital period distribution of binary stars having periods $<$10,000 years and primary star $T_{eff} \lesssim 6000$ K is shown.  Filled blue circles represent K and M dwarf primary stars, and is based on data compiled in literature \citep{hal03,mac04,bur07,bec08,ben08,ruc08,dim10,dav11} and at http://www.vlmbinaries.org.  Amongst these low mass stars, a distinct lack of low $q$ binaries at 
short orbital periods is present.  A similar trend is also observed in the distribution of confirmed binaries around solar-like (0.8 M$_{\Sun}$ $<$ M $<$ 1.1 M$_{\Sun}$) stars 
(filled green squares), as seen in data repoduced from tabulated data in literature \citep{hal03,ben08,rag10,sah11}.  The mass ratio of the TYC 4110 (large red cross and an arrow of arbitrary size), OGLE-TR-122 (small black cross; adopted from \citealt{pon05}), Kepler-16 (small black cross; adopted from \citealt{doy11}), and vB 69 (small black cross; adopted from \citealt{ben08}) systems are unique in that they lie within this mass ratio-period deficit.  \label{q}}
\end{center}
\end{figure}

Arguably the most distinctive 
feature of TYC 4110 is its extremely low mass ratio, $q$, of $\geq$0.087 $\pm$0.003, given its relatively short $\sim$79-day period.  As illustrated in Figure \ref{q}, previous statistical investigations of binarity 
in solar-like ($T_{eff} \lesssim 6000$ K) stars have found evidence that the 
short period brown dwarf desert extends in mass toward the low mass star regime \citep{bur07,bou11,sah11}.  A similar desert of short-period, low $q$ companions to low-mass K and M dwarf stars has also been observed (Figure \ref{q}; see also \citealt{bur07}).  The studies of companions around solar-like stars by both \citet{duq91} and \citet{rag10} report no firm detection of low $q$ binary 
companions with orbital periods $<$100 days.  The closest short period analog observed by \citet{rag10} (see e.g. their Figure 17) is a multiple ($>$ 2 components) $q$ $\sim$0.23 system with an orbital period of $\sim$50 days, while the closest short period binary is a $\sim$1-day period object with a $q$ of $\sim$0.4.  
All other low $q$ binaries in their sample have orbital periods $\geq$5000 days.  \citet{may01} also report a sparse number of low $q$ binary companions to 
solar-like stars, but do not quantify the orbital periods of these objects.  

The absolute ``dryness'' of the low mass ratio, short period desert for solar-like ($T_{eff} \lesssim 6000$ K) illustrated in Figure \ref{q} is a subject of active investigation and debate in the literature.  OGLE-TR-122b \citep{pon05}, Kepler-16b \citep{doy11}, and vB 69 \citep{ben08} are short period, low $q$, solar-like binary 
systems that populate this parameter space.  The statistical frequency of objects like OGLE-TR-122b and Kepler-16b have not been quantified by the OGLE or Kepler surveys; hence, it is plausible that they are indeed rare.  We note that there have been suggestions the mass ratios of binary companions surrounding F7 to K-type
stars are more broadly distrubuted and that no clear low mass ratio desert exists (see Figure 8 in \citealt{hal03}).  However, we have included all of the companions surrounding 
G-type (0.8 M$_{\Sun}$ $<$ M $<$ 1.1 M$_{\Sun}$) and K- and M-type primaries tabulated in \citet{hal03} (but omitting data from their survey which was not tabulated 
in print or online) in Figure \ref{q}, and still clearly see a short period, low mass ratio companion deficit.  Clearly, additional observational studies which precisely 
characterize the stellar properties of the host star and characterize the properties of their binary companions are needed to perform an accurate meta-analysis of the dependence of companion mass ratios as a function of orbital period and host star mass.  These studies should make careful note of their observational biases to facilitate 
a more robust cross-correlation of results compiled from different surveys.  A future meta-analysis would also particularly benefit from having current and future studies 
fully tabulating the fundamental properties of the binary systems they investigate.

As noted in Section \ref{intro}, short-period, low $q$ companions have been reported around stars slightly more massive that the Sun (F-type stars; \citealt{bou05,pon06,bea07,bou11}).  \citet{bou11} propose that short-period low $q$ companions might form around a wide mass range of stars, including G-type stars, 
but suggest that weaker magnetic disk braking during the early formation history of F-type stars might transfer less angular momentum to their companion bodies, thereby 
preventing catastrophic decays of their orbits.  Conversely, \citet{bou11} propose that a stronger disk braking in young G-type stars might distribute more angular momentum to 
their companion bodies, causing (initially) short-period companions to migrate inwards and become engulfed by the primary.  While intriguing, this proposed 
evolutionary scenario would clearly benefit by a more robust assessment of how ``dry'' the short-period, low $q$ desert is around GKM dwarfs as compared to F dwarfs.

The precise mass ratio of MARVELS-3B is not known due to the unknown inclination of the system, thus our observations only set a lower limit of $q$ of $>$0.087 $\pm$0.003.  
However, as demonstrated in Section \ref{companion}, our Bayesian analysis of the system using 4 plausible prior assumptions all indicate the likely median mass of 
MARVELS-3B is a dM star, yielding mass ratios of 0.087 $<$ $q$ $<$ 0.149 (Figure \ref{mcmc} and Table \ref{tableofq}).  Like OGLE-TR-122b \citep{pon05} and Kepler-16b \citep{doy11}, 
MARVELS-3B therefore seems likely to be an 
outlier to the mass ratio-period relationships commonly observed for both solar-like stars and lower-mass dM stars (Figure \ref{q}).  

In this context, we note that analyses of the period distribution of exoplanets have revealed a deficit of such bodies having orbital periods of 10-100 days, aka 
the ``period valley'' \citep{udr03,jon03}.  \citet{wit10} suggest that this observed deficit is real for the giant planet population (M $>$ 100 M$_{\Earth}$), but that 
any deficit of lower mass planets (10-100 M$_{\Earth}$) in this period regime might be the result of selection effects.  One possible explanation for the dearth of giant 
planets in 10-100 day orbits is a decrease in the amount of orbital migration which such objects experience (see e.g. \citealt{tri98}).  

Migration could also play a role in 
setting the observed period distribution of the low-mass stellar binary regime.  The orbital evolution of binaries has been explored computationally, and stellar accretion, the interaction between binaries with their natal gas disks, and interactions between triple components can influence these systems (see e.g. \citealt{bat02,sta09,kra11}).  
\citet{bat02}, for example, found such processes were successful at producing short-period binaries in high mass ratio systems ($q$ $\geq$ 0.3), but it is uncertain 
as to whether these specific simulations could produce short-period, low $q$ systems like MARVELS-3B.

N-body simulations of the early evolution of stellar clusters which include instantaneous gas removal \citep{moe10} are beginning to better reproduce the number 
of unequal mass solar-like binaries observed \citep{duq91,rag10}.  Potentially relevant for MARVELS-3B, \citet{moe10} showed that one simulated triple system comprised 
of a 0.96 and 0.73 M$_{\Sun}$, 2 AU separation binary with a 0.21 M$_{\Sun}$ tertiary at 12 AU, broke up into a tight (0.1 AU separation) binary system comprised of the 
0.21 and 0.73 M$_{\Sun}$ components (e.g. a mass ratio, $q$, of 0.29).  We therefore speculate that it is possible that the short period, low $q$ 
MARVELS-3B binary was initially part of a tertiary (or larger) system with much different initial orbital parameters, and only achieved its final orbital configuration 
following the dispersal of the cluster in which it formed.  Although our analysis of available SuperWASP photometry and MARVELS+SARG radial velocity data exhibited no evidence of the presence of additional, long-period ($\geq$ 2 years) bodies in the system, our single-epoch detection of a candidate tertiary body in the system via high contrast imaging could support this interpretation.  Additional epochs of imagery should be pursued to establish if this body is co-moving with the TYC 4110 system.  Although beyond the scope of this paper, future 
simulations of binary migration and orbital evolution during cluster dispersal should explore the limits and frequency at 
which they can reproduce short period, low mass ratio binaries for solar-like stars.

\section{Conclusions}

We present a detailed analysis of the fundamental properties of the solar-like star TYC 4110-01037-1 and its very low mass stellar companion.  This analysis was performed 
in the context of our long-term goal of performing a detailed statistical analysis of global trends in the population of well vetted and characterized BDs and low mass binary companions identified by the MARVELS survey.  We find:

\begin{itemize}
\item TYC 4110-01037-1 is a $\lesssim$5 Gyr solar-like star characterized by $T_{eff}$ = 5879 $\pm$29 K, \hbox{$\log g$} = 4.48 $\pm$0.15,  and $[Fe/H]$ = -0.01 $\pm$0.05.  
We determine the stellar mass to be 1.07 $\pm$0.08 M$_{\Sun}$ and stellar radius to be 0.99 $\pm$0.18 R$_{\Sun}$. 

\item MARVELS-3B is a $>$97.7 $\pm$5.8 M$_{Jup}$ (M$\sin i$) companion to TYC 4110-01037-1, which follows a moderately elliptical (e = 0.1095 $\pm$0.0023), 
78.994 $\pm$0.012 day orbital period.  

\item The mass ratio, q, of the companion to the primary is $\geq$ 0.087 $\pm$0.003.  MARVELS-3B therefore resides in a 
short period, low $q$ desert analogous to the short period brown dwarf desert.

\item We speculate that MARVELS-3B might have initially formed in a tertiary system with much different orbital parameters, and achieved its present day configuration 
following the dispersal of the cluster in which it formed.  A candidate tertiary body has been identified via single-epoch, high contrast imagery.  If this object is confirmed to be co-moving, we estimate it would be a dM4 star.

\end{itemize}

\acknowledgements 
We thank the referee for providing feedback which helped to improve the content and clarity of this manuscript.  We also thank D. Raghavan for providing an electronic table of his published G dwarf binaries.  This research was partially supported by NSF AAPF AST 08-02230 (JPW), NSF CAREER Grant AST 0645416 (EA), the Vanderbilt Initiative in Data-Intensive Astrophysics (VIDA) and NSF CAREER Grant AST 0349075 (KGS,LH,JP), CNPq grant 476909/2006-6 (GFPM), FAPERJ grant APQ1/26/170.687/2004 (GFPM), and a 
PAPDRJ CAPES/FAPERJ  fellowship (LG).  Funding for the MARVELS multi-object Doppler instrument was provided by the W.M. Keck Foundation and NSF grant AST-0705139. The MARVELS survey was partially funded by the SDSS-III consortium, NSF Grant AST-0705139, NASA with grant NNX07AP14G and the University of Florida.  The Center for Exoplanets and Habitable Worlds is supported by the Pennsylvania State University, the Eberly College of Science, and the Pennsylvania Space Grant Consortium.

This work has made use of observations taken with the Telescopio Nationale Galileo (TNG) operated on the island of La Palma by the Fundation Galileo Galilei, funded by the Instituto Nazionale di Astrofisica (INAF), in the Spanish {\it Observatorio del Roque de los Muchachos} of the Instituto de Astrof{\'\i}sica de Canarias (IAC). 
We also make use of data products from the Wide-field Infrared Survey Explorer, which is a joint project of the University of California, Los Angeles, and the Jet Propulsion Laboratory/California Institute of Technology, funded by the National Aeronautics and Space Administration.  Funding for SDSS-III has been provided by the Alfred P. Sloan Foundation, the Participating Institutions, the National Science Foundation, and the U.S. Department of Energy Office of Science.  The SDSS-III web site is http://www.sdss3.org/.

SDSS-III is managed by the Astrophysical Research Consortium for the Participating Institutions of the SDSS-III Collaboration including the University of Arizona, the Brazilian Participation Group, Brookhaven National Laboratory, University of Cambridge, University of Florida, the French Participation Group, the German Participation Group, the Instituto de Astrofisica de Canarias, the Michigan State/Notre Dame/JINA Participation Group, Johns Hopkins University, Lawrence Berkeley National Laboratory, Max Planck Institute for Astrophysics, New Mexico State University, New York University, Ohio State University, Pennsylvania State University, University of Portsmouth, Princeton University, the Spanish Participation Group, University of Tokyo, University of Utah, Vanderbilt University, University of Virginia, University of Washington, and Yale University.

\newpage
\clearpage
\begin{table}
\begin{center}
\footnotesize
\caption{Summary of Observed Radial Velocities\label{rv}}
\begin{tabular}{lccc}
\tableline
HJD & Instrument & RV & $\sigma_{RV}$ \\
  & & km s$^{-1}$ & km s$^{-1}$  \\
\tableline
2454811.815473 & M & -2.869 & 0.053\\
2454812.936774 & M & -2.386 & 0.049\\
2454816.915562 & M & -0.841 & 0.060\\
2454840.865816 & M &  3.859 & 0.046\\
2454842.900506 & M &  3.614 & 0.093\\
2454843.814123 & M &  3.496 & 0.057\\
2454844.776592 & M &  3.342 & 0.045\\
2454845.779836 & M &  3.153 & 0.049\\
2454867.724621 & M & -2.229 & 0.072\\
2454868.791455 & M & -2.527 & 0.048\\
2454869.785190 & M & -2.786 & 0.070\\
2454901.671359 & M &  1.393 & 0.068\\
2455105.974063 & M & -2.454 & 0.046\\
2455135.969729 & M &  0.417 & 0.039\\
2455141.825627 & M &  2.508 & 0.045\\
2455142.892315 & M &  2.744 & 0.060\\
2455143.904775 & M &  2.978 & 0.058\\
2455144.904270 & M &  3.187 & 0.042\\
2455161.844226 & M &  3.115 & 0.047\\
2455199.808399 & M & -4.240 & 0.059\\
2455201.786093 & M & -4.002 & 0.040\\
2455202.835898 & M & -3.854 & 0.042\\
2455280.657591 & M & -4.064 & 0.046\\
2455287.644203 & M & -2.150 & 0.052\\
2455463.908559 & M &  3.816 & 0.073\\
2455470.923509 & M &  4.025 & 0.042\\
2455472.000492 & M &  3.942 & 0.047\\
2455498.003488 & M & -1.744 & 0.066\\
2455500.992667 & M & -2.521 & 0.059\\
2455516.486551 & S & -4.209 & 0.061\\
2455516.579785 & S & -4.118 & 0.033\\
2455542.731058 & M &  3.798 & 0.049\\
2455545.783002 & M &  4.117 & 0.053\\
2455553.524971 & S &  3.677 & 0.036\\
2455556.930997 & M &  3.175 & 0.076\\
2455580.495813 & S & -2.702 & 0.034\\
2455666.464340 & S & -4.051 & 0.037\\
2455698.384891 & S &  3.375 & 0.041\\
\tableline
\end{tabular}
\end{center}
\vspace{-0.3in}
\tablecomments{A summary of relative radial velocities obtained with the MARVELS (M) and SARG (S) spectrographs.  The quoted $\sigma_{RV}$ errors for the MARVELS data were first uniformly scaled by a ``quality factor'' Q =5.67 (see \citealt{fle10}), based on the rms of the other stars observed on the same 
SDSS-III plate.  As described in Section \ref{companion}, the errors of the MARVELS and SARG data were then re-scaled to force our RV fits to have P($\chi^{2}$) = 0.5, in an iterative 
manner.  Note that zero point offsets have been applied to the RVs compiled in column 3 to force the RVs to vary about 0 m s$^{-1}$.}
\end{table}

\newpage
\clearpage
\begin{table}
\begin{center}
\footnotesize
\caption{Stellar Properties of TYC 4110-01037-1 \label{star}}
\begin{tabular}{lccc}
\tableline
Parameter & Value & Uncertainty & Note\\
\tableline
$\alpha$ (2000) & 06 54 12.3 & \nodata & \citet{hog98} \\
$\delta$ (2000) & +60 21 31.4 & \nodata & \citet{hog98} \\
NUV & 15.715 mag & 0.016 & GALEX \citep{mor07} \\
B & 11.159 mag & 0.018 & this work (HAO) \\
V & 10.521 mag & 0.019 & this work (HAO) \\
Ic & 9.788 mag & 0.037 & this work (HAO) \\
J & 9.347 mag & 0.023 &  \citet{cut03} \\
H & 9.069 mag & 0.210 &  \citet{cut03} \\
Ks & 9.004 mag & 0.020 &  \citet{cut03} \\
WISE1 (3.4 $\mu$m) & 8.932 mag & 0.024 & \nodata \\
WISE2 (4.6 $\mu$m) & 8.970 mag & 0.022 & \nodata \\
WISE3 (12 $\mu$m) & 8.934 mag & 0.030 & \nodata \\
WISE4 (22 $\mu$m) &  8.662 mag & 0.344 & \nodata \\
A$_{V}$ & 0.16 & 0.04 & this work \\
d & 125.1 pc & 4.6 & this work \\
Teff & 5879 K & 29  & this work \\
$\log g$ [cgs] & 4.48 & 0.15 & this work \\
$[Fe/H]$ & -0.01 & 0.05 & this work \\
v$_{micro}$ & 1.40 km s$^{-1}$ & 0.05  & this work  \\
M$_{\star}$ & 1.07 M$_{\Sun}$ & 0.08 & this work \\
R$_{\star}$ & 0.99 R$_{\Sun}$ & 0.18 & this work \\
v$_{systemic}$ & 25.9 km s$^{-1}$ & 0.2 & \nodata \\
v$_{rot} sin i$ & $\lesssim$3 km s$^{-1}$ & \nodata & \nodata \\
\tableline
\end{tabular}
\end{center}
\vspace{-0.3in}
\tablecomments{A summary of some of the basic observational properties of TYC 4110-01037-1.  }
\end{table}

\newpage
\clearpage
\begin{table}
\begin{center}
\footnotesize
\caption{Properties of MARVELS-3B \label{fit}}
\begin{tabular}{lc}
\tableline
Parameter & Value \\
\tableline
$T_{C}$ \ (BJD$_{TDB}$ - 2450000) & $5175.97\pm0.14$\\
                 $P$\ (days) & $78.994\pm0.012$\\
                         $e$ & $0.1095\pm0.0023$\\
         $\omega$\ (radians) & $4.380_{-0.042}^{+0.041}$\\
                  $K$\ (m/s) & $4199\pm11$\\
       $\gamma_{TNG}$\ (m/s) & $1338\pm19$\\
       $\gamma_{APO}$\ (m/s) & $2945.0_{-9.9}^{+10.0}$\\
              ecos$(\omega)$ & $-0.0357_{-0.0043}^{+0.0042}$\\
              esin$(\omega)$ & $-0.1034\pm0.0027$\\
$T_{P}$ \ (BJD$_{TDB}$ - 2450000) & $5210.32_{-0.51}^{+0.50}$\\
a (sin i = 1) & 0.38 AU \\
M (M$_{Jup}$) & $>$97.7$\pm$5.8 \\
\tableline
\end{tabular}
\end{center}
\vspace{-0.3in}
\tablecomments{A summary of some of the basic orbital properties of MARVELS-3B.  The quoted minimum mass corresponds to the limiting case of $\sin i$ = 1.  
Note that T$_{P}$ corresponds to the time of periastron while T$_{C}$ 
corresponds to the time of conjunction.}
\end{table}

\newpage
\clearpage
\begin{table}
\begin{center}
\footnotesize
\caption{MCMC Properties of MARVELS-3B \label{tableofq}}
\begin{tabular}{lccc}
\tableline
Assumed Prior & Mass of MARVELS-3B & Transit Probability & q (Mass Ratio) \\
\tableline
none ($\sin i$ = 1) & 97.7 M$_{Jup}$ & 1 & $>$0.087 $\pm$0.003 \\
$dN/dq$ $\propto$ q$^{+1}$ & 166.2 M$_{Jup}$ & 0.0055 & 0.149 \\
$dN/dq$ $\propto$ q$^{-1}$ & 113.0 M$_{Jup}$ & 0.0129 & 0.100 \\
$dN/dq$ = const & 125.9 M$_{Jup}$ & 0.0092 & 0.112  \\
\tableline
\end{tabular}
\end{center}
\vspace{-0.3in}
\tablecomments{A summary different mass estimates of MARVELS-3B and the resultant mass ratio for different assumed priors on the statistical 
companion mass ratio around G dwarfs and our MCMC analysis.  The prior ``none'' corresponds to the minimum mass of MARVELS-3B, simply 
assuming $\sin i$ = 1.  For all other priors, the quoted companion mass and q values are the median mass determined from our MCMC analysis described in 
Section \ref{companion} and shown in Figure \ref{mcmc}.}
\end{table}

\newpage
\clearpage
\begin{table}
\begin{center}
\footnotesize
\caption{Properties of the Candidate Tertiary Companion to TYC 4110-01037-1 \label{3table}}
\begin{tabular}{lc}
\tableline
Filter & Magnitude \\
\tableline
J  & 13.59 $\pm$ 0.13 \\
H &  13.04 $\pm$ 0.24 \\
K$^{'}$ & 12.84 $\pm$ 0.05 \\
\tableline
\end{tabular}
\end{center}
\vspace{-0.3in}
\tablecomments{The computed apparent magnitude of the candidate tertiary companion to TYC 4110 detected in Figure \ref{ao} is compiled.}
\end{table}

\newpage
\clearpage

\newpage
\clearpage

\newpage
\clearpage

\newpage
\clearpage

\newpage
\clearpage

\newpage
\clearpage

\newpage
\clearpage

\newpage
\clearpage

\newpage
\clearpage


\begin{thebibliography}{}
\bibitem[Asplund et al.(2009)]{asp09} Asplund, M., Grevesse, N., Sauval, A.~J., \& Scott, P.\ 2009, \araa, 47, 481
\bibitem[Baluev(2008)]{bal08} Baluev, R.V. 2008, MNRAS, 385, 1279 
\bibitem[Baraffe et al.(1998)]{bar98} Baraffe, I., Chabrier, G., Allard, F., \& Hauschildt, P.H. 1998, A\&A, 337, 403
\bibitem[Bate et al.(2002)]{bat02} Bate, M.R., Bonnell, I.A., \& Bromm, V. 2002, MNRAS, 336, 705
\bibitem[Beatty et al.(2007)]{bea07} Beatty, T.G. et al. 2007, ApJ, 663, 573
\bibitem[Becker et al.(2008)]{bec08} Becker, A.C. et al. 2008, MNRAS, 386, 416
\bibitem[Bender \& Simon(2008)]{ben08} Bender, C.F. \& Simon, M. 2008, ApJ, 689, 416
\bibitem[Biller et al.(2010)]{bil10} Biller, B.A. et al. 2010, ApJL, 720, 82
\bibitem[Borucki et al.(2011a)]{bo11a} Borucki, W.J. et al. 2011a, ApJ, 728, 117
\bibitem[Borucki et al.(2011b)]{bo11b} Borucki, W.J. et al. 2011b, ApJ, 736, 19
\bibitem[Bouchy et al.(2005)]{bou05} Bouchy, F., Pont, F., Melo, C., Santos, N.C., Mayor, M., Queloz, D., \& Udry, S. 2005, A\&A, 431, 1105
\bibitem[Bouchy et al.(2011)]{bou11} Bouchy, F. et al. 2011, A\&A, 533, 83
\bibitem[Burgasser et al.(2007)]{bur07} Burgasser, A.J., Reid, I.N., Siegler, N., Close, L., Allen, P., Lowrance, P., \& Gizis, J. 2007, Protostars and Planets V, ed. B. Reipurth, D. Jewitt, \& K. Keil (Tucson, AZ: Univ. Arizona Press), 427
\bibitem[Butters et al.(2010)]{but10} Butters, O.W. et al. 2010, A\&A, 520, L10
\bibitem[Castelli \& Kurucz(2004)]{cas04} Castelli, F. \& Kurucz, R. 2004, astro-ph/0405087
\bibitem[Cox(2000)]{cox00} Cox, A.N. 2000, Allens's Astrophysical Quantities, 4th ed. (New York: AIP Press)
\bibitem[Crepp et al.(2012)]{cre12} Crepp, J.R. et al 2012, ApJ, submitted (astro-ph/1112.1725)
\bibitem[Cumming(2004)]{cum04} Cumming, A. 2004, MNRAS, 354, 1165
\bibitem[Cutri et al.(2003)]{cut03} Cutri, R.M. et al. 2003, 2MASS All-Sky Catalog of Point Sources, VizieR Online Data Catalog II/246
\bibitem[Davenport et al.(2011)]{dav11} Davenport, J.R.A. et al. 2011, ApJL, in prep
\bibitem[Demarque et al.(2004)]{dem04} Demarque, P., Woo, J., Kim, Y., \& Yi, S.K. 2004, ApJS, 155, 667
\bibitem[Dimitrov \& Kjurkchieva(2010)]{dim10} Dimitrov, D.P. \& Kjurkchieva, D.P. 2010, MNRAS, 406, 2559
\bibitem[Doyle et al.(2011)]{doy11} Doyle, L.R. et al. 2011, Science, 333, 1602
\bibitem[Duquennoy \& Mayor(1991)]{duq91} Duquennoy, A. \& Mayor, M. 1991, A\&A, 248, 485
\bibitem[Eastman et al.(2012)]{eas11} Eastman, J. et al. 2012, in prep
\bibitem[Eisenstein et al.(2011)]{eis11} Eisenstein, D.J. et al. 2011, AJ, 142, 72
\bibitem[Fleming et al.(2010)]{fle10} Fleming, S.W. et al. 2010, ApJ, 718, 1186
\bibitem[Fleming et al.(2012)]{fle11} Fleming, S.W. et al. 2012, AJ, in prep 
\bibitem[Ge et al.(2009)]{ge09} Ge, J. et al. 2009, Proc. SPIE, 7440, 74400L
\bibitem[Girardi et al.(2002)]{gir02} Girardi, L. et al. 2002, A\&A, 391, 195
\bibitem[Ghez et al.(2008)]{ghe08} Ghez, A.M. et al. 2008, ApJ, 689, 1044
\bibitem[Gizis et al.(2001)]{giz01} Gizis, J.E., Kirkpatrick, J.D., Burgasser, A., Reid, I.N., Monet, D.G., Liebert, J., \& Wilson, J.C. 2001, ApJ, 551, L163
\bibitem[Gratton et al.(2001)]{gra01} Gratton, R.G. et al. 2001, Exp. Astron., 12, 107
\bibitem[Grether \& Lineweaver(2006)]{gre06} Grether, D. \& Lineweaver, C.H. 2006, ApJ, 640, 1051
\bibitem[Gunn et al.(2006)]{gun06} Gunn, J.E. et al. 2006, AJ, 131, 2332
\bibitem[Halbwachs et al.(2003)]{hal03} Halbwachs, J.L., Mayor, M., Udry, S., \& Arenou, F. 2003, A\&A, 397, 159
\bibitem[Hauschildt et al.(1999)]{hau99} Hauschildt, P.H., Allard, F., \& Baron, E. 1999, ApJ, 512, 377
\bibitem[Hog et al.(1998)]{hog98} Hog, E. et al. 1998, A\&A, 335, 65
\bibitem[Holweger et al. (1991)]{hol91} Holweger, H., Bard, A., Kock, M., \& Kock, A.\ 1991, \aap, 249, 545
\bibitem[Janson et al.(2011)]{jan11} Janson, M. et al. 2011, ApJL, 728, 85
\bibitem[Jones et al.(2003)]{jon03} Jones, H.R.A., Butler, R.P., Tinney, C.G., Marcy, G.W., Penny, A.J., McCarthy, C., \& Carter, B.D. 2003, MNRAS, 341, 948
\bibitem[Kalas et al.(2008)]{kal08} Kalas, P. et al. 2008, Science, 322, 1345
\bibitem[Kratter(2011)]{kra11} Kratter, K.M. 2011, ASP Conf Ser, Evolution of Compact Binaries, ed. L. Schmidtobreick, M.R. Schreiber, \& C. Tappert, in press (astro-ph/1109.3740)
\bibitem[Kraus \& Hillenbrand(2007)]{kra07} Kraus, A.L. \& Hillenbrand, L.A. 2007, AJ, 134, 2340
\bibitem[Kraus et al.(2011)]{kra11} Kraus, A.L., Ireland, M.J., Martinache, F., \& Hillenbrand, L.A. 2011, ApJ, 731, 8
\bibitem[Kupka et al.(1999)]{kup99} Kupka, F., Piskunov, N., Ryabchikova, T.~A., Stempels, H.~C., \& Weiss, W.~W.\ 1999, \aaps, 138, 119
\bibitem[Kurucz et al.(1984)]{kur84} Kurucz, R.~L., Furenlid, I., Brault, J., \& Testerman, L.\ 1984, National Solar Observatory Atlas, Sunspot, New Mexico: National Solar Observatory, 1984
\bibitem[Kurucz(1993)]{kur93} Kurucz, R. 1993, ATLAS9 Stellar Atmosphere Programs and 2 km s$^{-1}$ grid. Kurucz CD-ROM No. 13, Cambridge, MA: SAO, 1993, 13
\bibitem[Lagrange et al.(2010)]{lag10} Lagrange, A.-M. 2010, Science, 329, 57
\bibitem[Landolt(1992)]{lan92} Landolt, A.U. 1992, AJ, 104, 340
\bibitem[Lee et al.(2011)]{lee11} Lee, B.L. 2011, ApJ, 728, 32
\bibitem[Leggett et al.(2002)]{leg02} Leggett, S.K. et al. 2002, ApJ, 564, 452
\bibitem[Lomb(1976)]{lom76} Lomb, N.R. 1976, ApSS, 39, 447
\bibitem[Maceroni \& Montalban(2004)]{mac04} Maceroni, C. \& Montalban, J. 2004, A\&A, 426, 577 
\bibitem[Marcy \& Butler(2000)]{mar00} Marcy, G.W. \& Butler, R.P. 1990, PASP, 112, 137
\bibitem[Marois et al.(2008)]{mar08} Marois, C. et al. 2008, Science, 322, 1348
\bibitem[Marois et al.(2010)]{mar10} Marois, C., Zuckerman, B., Konopacky, Q.M., Macintosh, B., \& Barman, T. 2010, Nature, 468, 1080
\bibitem[McCarthy \& Zuckerman(2004)]{mcc04} McCarthy, C. \& Zuckerman, B. 2004, AJ, 127, 2871
\bibitem[Mayor et al.(1992)]{may92} Mayor, M., Duquennoy, A., Halbwachs, J., \& Mermilliod, J. 1992, ASP Conf Ser. 32, IAU Colloq. 135, Complementary Approaches to Double and Multiple Star Research, ed. H.A. McAlister \& W.I. Hartkopf (San Francisco, CA:ASP), 73
\bibitem[Mayor et al.(2001)]{may01} Mayor, M. et al. 2001, in IAU Symp 200, The Formation of Binary Stars, ed. B. Reipurth \& H. Zinnecker (San Francisco: ASP), 93
\bibitem[Mayor et al.(2004)]{may04} Mayor, M., Udry, S., Naef, D., Pepe, F., Queloz, D., Santos, N.C., \& Burnet, M. 2004, A\&A, 415, 391
\bibitem[Metchev \& Hillenbrand(2004)]{met04} Metchev, S. A. \& Hillenbrand, L.A. 2004, ApJ, 617, 1330
\bibitem[Metchev \& Hillenbrand(2009)]{met09} Metchev, S.A. \& Hillenbrand, L.A. 2009, ApJS, 181, 62
\bibitem[Moeckel \& Bate(2010)]{moe10} Moeckel, N. \& Bate, M.R. 2010, MNRAS, 404, 721
\bibitem[Moore et al.(1966)]{mo66} Moore, C.~E., Minnaert, M.~G.~J., \& Houtgast, J.\ 1966, National Bureau of Standards Monograph, Washington: US Government Printing Office (USGPO), 1966
\bibitem[Morrissey et al.(2007)]{mor07} Morrissey, P. et al. 2007, ApJS, 173, 682
\bibitem[Muirhead et al.(2011)]{mui11} Muirhead, P.S., Hamren, K., Schlawin, E., Rojas-Ayala, B., Covey, K.R., \& Lloyd, J.P. 2011, ApJL, submitted (astro-ph/1109.1819)
\bibitem[Patel et al.(2007)]{pat07} Patel, S., Vogt, S.S., Marcy, G.W., Johnson, J.A., Fischer, D.A., Wright, J.T., \& Butler, R.P. 2007, ApJ, 665, 744
\bibitem[Pickles(1998)]{pic98} Pickles, A.J. 1998, PASP, 110, 863
\bibitem[Pollacco et al.(2006)]{pol06} Pollacco, D.L. et al. 2006, PASP, 118, 1407
\bibitem[Pont et al.(2005)]{pon05} Pont, F., Melo, C.H.F., Bouchy, F., Udry, S., Queloz, D., Mayor, M., \& Santos, N.C. 2005, A\&A, 433, L21
\bibitem[Pont et al.(2006)]{pon06} Pont, F. et al. 2006, A\&A, 447, 1035
\bibitem[Press \& Rybicki(1989)]{pre89} Press, W.H., \& Rybicki, G.B. 1989, ApJ, 338, 277 
\bibitem[Press et al.(1992)]{pre92} Press, W.H., Teukolsky, S.A., Vetterling, W.T., \& Flannery, B.P. 1992, Cambridge: University Press, |c1992, 2nd ed.
\bibitem[Raghavan et al.(2010)]{rag10} Raghavan, D. et al. 2010, ApJS, 190, 1
\bibitem[Rucinski \& Pribulla(2008)]{ruc08} Rucinski, S.M. \& Pribulla, T. 2008, MNRAS, 388, 1831
\bibitem[Sahlmann et al.(2011)]{sah11} Sahlmann, J. et al. 2011, A\&A, 525, 95
\bibitem[Scargle(1982)]{sca82} Scargle, J.D. 1982, ApJ, 263, 835
\bibitem[Schlegel et al.(1998)]{sch98} Schlegel, D.J., Finkbeiner, D.P., \& Davis, M. 1998, ApJ, 500, 525
\bibitem[Sneden(1973)]{sne73} Sneden, C. 1973, PhD Thesis, Univ. of Texas-Austin
\bibitem[Sousa et al. (2007)]{sou07} Sousa, S.~G., Santos, N.~C., Israelian, G., Mayor, M., \& Monteiro, M.~J.~P.~F.~G.\ 2007, \aap, 469, 783
\bibitem[Sousa et al.(2008)]{sou08} Sousa, G. et al. 2008, A\&A, 487, 373
\bibitem[Stamatellos \& Whitworth(2009)]{sta09} Stamatellos, D. \& Whitworth, A.P. 2009, MNRAS, 392, 413
\bibitem[Tabernero et al.(2012)]{tab11} Tabernero, H., Montes, D., \& Gonzalez Hernandez, J.I. 2012, A\&A, submitted
\bibitem[Thalmann et al.(2009)]{tha09} Thalmann, C. et al. 2009, ApJL, 707, 123
\bibitem[Tinney et al.(2001)]{tin01} Tinney, C.G. et al. 2001, ApJ, 551, 507
\bibitem[Tokovinin(2004)]{tok04} Tokovinin, A. 2004, RevMexAA Conf. Ser., 21, 7
\bibitem[Torres et al.(2010)]{tor10} Torres, G., Andersen, J., \& Gimenez, A. 2010, A\&A Rev., 18, 67
\bibitem[Trilling et al.(1998)]{tri98} Trilling, D.E., Benz, W., Guillot, T., Lunine, J.I., Hubbard, W.B., \& Burrows, A. 1998, ApJ, 500, 428
\bibitem[Udry et al.(2000)]{udr00} Udry, S., et al. 2000, A\&A, 356, 590
\bibitem[Udry et al.(2003)]{udr03} Udry, S., Mayor, M., \& Santos, N.C. 2003, A\&A, 407, 369
\bibitem[Vogt et al.(2002)]{vog02} Vogt, S.S., Butler, R.P., Marcy, G.W., Fischer, D.A., Pourbaix, D., Apps, K., \& Laughlin, G. 2002, ApJ, 568, 352
\bibitem[Wahhaj et al.(2011)]{wah11} Wahhaj, Z. et al. 2011, ApJ, 729, 139
\bibitem[Wang et al.(2003)]{wan03} Wang, S. et al. 2003, Proc. SPIE, 4841, 1145
\bibitem[Wittenmyer et al.(2010)]{wit10} Wittenmyer, R.A., OToole, S.J., Jones, H.R.A., Tinney, C.G., Butler, R.P., Carter, B.D., \& Bailey, J. 2010, ApJ, 722, 1854
\bibitem[Wright et al.(2010)]{wri10} Wright, E.L. et al. 2010, AJ, 140, 1868
\end{thebibliography}
\end{document}